\documentclass[a4paper,11pt,
]{article}
\usepackage{jcappub}
\usepackage{booktabs}
\usepackage{float}
\usepackage[dvipsnames]{xcolor}
\usepackage[normalem]{ulem}
\usepackage{cleveref}
\usepackage{csquotes}

\DeclareMathOperator{\Tr}{Tr}

\title{A speed limit on tachyon fields from cosmological and fine-structure data}

\author[1,2,3]{J. D. F. Dias,}%
\affiliation[1]{Centro de Astrof\'{\i}sica da Universidade do Porto,\\Rua das Estrelas, 4150-762 Porto, Portugal}
\affiliation[2]{Instituto de Astrof\'isica e Ci\^encias do Espa\c co, CAUP, Universidade do Porto, Rua das Estrelas, 4150-762, Porto, Portugal}
\affiliation[3]{Faculdade de Ci\^encias, Universidade do Porto, Rua Campo Alegre, 4169-007, Porto, Portugal}
\emailAdd{joao.dias@astro.up.pt}
\author[4]{Nils Sch\"oneberg,}%
\affiliation[4]{Institut de Ci\`encies del Cosmos Universitat de Barcelona (IEEC-UB),\\ Mart\'i i Franqu\'es, 1, E08028 Barcelona, Spain}%
\emailAdd{nils.science@gmail.com}
\author[5]{L\'eo Vacher,}
\affiliation[5]{Institute for Astrophysics and Cosmology, SISSA,\\ Via Bonomea 265, 34136, Trieste, Italy}
\emailAdd{lvacher@sissa.it}
\author[1,2]{C. J. A. P. Martins,}
\emailAdd{Carlos.Martins@astro.up.pt}
\author[6]{Samy Vinzl}
\affiliation[6]{Universit\'e Toulouse III - Paul Sabatier (UPS),\\ 118 Route de Narbonne, 31062 Toulouse CEDEX 9, France}
\emailAdd{samy.vinzl@univ-tlse3.fr}

\abstract{The rolling tachyon is a non-canonical scalar field model  well motivated in string theory which naturally predicts variations of the fine-structure constant. Such variations can in principle lead to interesting observable consequences, but they can also lead to extremely tight constraints on these kinds of models. In this work we subject the rolling tachyon model evolving in a variety of potentials to current data and show that most cosmologically interesting evolutions are already strongly excluded. We find $|1+w_0| < 10^{-3}$ from cosmological data and $|1+w_0| < 10^{-9}$ from fine-structure data, leaving the rolling tachyon to either play a role almost entirely equivalent to a cosmological constant or that of a test field. We also find that in most of the allowed parameter space the field evolves very slowly, allowing its evolution to be approximated as an equivalent canonical scalar field.}

\begin{document}
\maketitle
\flushbottom

\section{Introduction}
\label{sec:intro}

The discovery of the late-time acceleration of the universe was first suggested by supernovae observations \cite{Riess1998,Perlmutter1999} and subsequently supported by multiple independent datasets \cite{Jimenez2002,Eisenstein2005,Betoule2014,Haridasu2017,PlanckVI2020}. 
The source for this phenomenon was dubbed dark energy, and despite this discovery not being recent, there is yet no clear scientific agreement about its origin. The $\Lambda$CDM model incorporates the simplest explanation: the acceleration originates from a cosmological constant, $\Lambda$, naturally present within general relativity. However, there is a disparity between the energy scale required for the cosmic acceleration and that predicted by particle physics \cite{Weinberg1989}. This problem can be addressed either by accepting that dark energy is governed by a cosmological constant and finding a mechanism responsible for the small value of $\Lambda$, or by finding an alternative way to describe dark energy. The second method can be approached in two ways, the first is by modifying the right-hand side of the Einstein equations, adding new physical entities which contribute to the total energy-momentum tensor $T_{\mu\nu}$ with a negative pressure -- this is the case for the quintessence \cite{Tsujikawa_2013}. The second approach is to modify the left-hand side of the Einstein equations, which is the case for modified gravity theories such as the well-known $f\left(R\right)$ gravity \cite{fr}. However, these two approaches are often equivalent under certain limits, such that modifying the gravity Lagrangian can be understood as the addition of new fields \cite{Velasquez2018}.

In this work, we study an alternative to the cosmological constant provided by non canonical scalar fields --- rolling tachyons --- as candidates to describe the recent acceleration of the universe \cite{Sen2002a, Sen2002b, Padmanabhan2002}. A rolling tachyon is an example of a Born-Infeld scalar, and these are well motivated in string theory. The tachyon Lagrangian is analogous to the Lagrangian of a relativistic particle, much in the same way as the quintessence one is to that of a non-relativistic particle, with the interesting characteristic that the interaction of scalar fields with gauge fields (as well as the dynamics of these fields) naturally leads to fine-structure constant variations. In contrast to many phenomenological quintessence fields models, where the couplings to matter and radiation have to be postulated (such as for example the Bekenstein model \cite{bekensteinOriginal}), in the rolling tachyon these can be obtained directly from a fundamental theory, specifically from an effective $D$-brane action, the Dirac-Born-Infeld (DBI) action \cite{Leigh1989}. 

In the context of the rolling tachyon one can choose a number of potentials that can reproduce the desired late-time cosmic acceleration \cite{Copeland2005}. In fact, one can find a potential to emulate virtually any desired cosmological evolution\footnote{This is akin to what happens for quintessence-type models but, as we will show, there are broad differences between the two classes of models, which lead to significantly different constraints for the two classes.}. This fact imprints a degeneracy in this description and, to alleviate this, one has to pick out physical criteria to further constrain the model. A good physical criterion often used to break these degeneracies is the variation of the fine-structure constant. Indeed, due to the presence of a coupling between the tachyon field and the electromagnetic sector in the DBI action, a time dependence of the fine-structure constant is an unavoidable prediction of such models.

Fundamental constants of nature appear everywhere in our physical theories and are quantities that, by definition, are not explained by the theory. So, these constants have to be directly measured, in order to allow for any quantitative prediction of the theory. Dirac was a pioneer in questioning the constancy of such parameters \cite{Dirac1937, Dirac1938}, which opened a novel path of research. It was first proposed by Jordan, in \cite{Jordan1937}, that these fundamental constants of nature could become dynamical fields and this now appears as a prediction in multiple frameworks beyond the standard model, such as string theory. Today, we have an extensive literature on variation of fundamental constants, in particular of the fine-structure constant, setting very tight constraints on these models, for reviews see for example \cite{Uzan2011,Martins2017}. As such, we expect rolling tachyon scenarios to be similarly constrained by these datasets. 

This work builds upon previously conducted studies \cite{Martins2016,Martins2021}, improving constraints and providing with a more systematic study of different potentials. In \cref{sec:gen}, we present the model and its more general features, in addition to the data that we used to calculate the constraints. In \cref{sec:models}, we analyse the potentials individually, describing the dynamics of the field and presenting the contour plots. Finally, we provide a more general discussion of the results and present our conclusions in \cref{sec:concl}.

\section{General aspects}
\label{sec:gen}
\subsection{Theoretical background}
\label{sec:theory}

The D$p$-branes are the $p$-dimensional spaces in which the endpoints of open strings are allowed to move, assuming Dirichlet boundary conditions. They are, by definition, omnipresent in any formulation of string theory and can be treated as independent objects, with momentum, energy, and charges, playing a dynamical role as important as the strings themselves. The low energy effective action describing the dynamics of D-branes is known as the Dirac-Born-Infeld action \cite{Leigh1989}, originally proposed as an extension of electromagnetism \cite{Dirac1962,Born-Infeld1934}. This action contains massless and massive scalar fields as well as gauge fields and a tachyon, which we will refer to as the rolling tachyon \cite{Sen2002a}. Tachyons can be identified by the presence of an imaginary mass, characteristic of what would be interpreted classically as resulting in superluminal motion. However, their corresponding quantum fields still obey causal commutation relations, and therefore preserve causality.

An important feature of the tachyon field, cosmologically speaking, is its non-canonical Lagrangian, which results in perhaps unexpected expressions for the energy density and pressure of the field. Indeed, the Lagrangian density for a homogeneous tachyon field is
\begin{equation} \label{eq:lagrangian}
    \mathcal{L} =-V\left(\phi\right)\sqrt{1-\dot{\phi}^2}~,
\end{equation}
which leads to the following expressions for energy density and pressure
\begin{equation}\label{eq:rho_P}
    \rho_{\phi}=\frac{V\left(\phi\right)}{\sqrt{1-\dot{\phi}^2}}~, \qquad \qquad
    P_{\phi}=-V\left(\phi\right)\sqrt{1-\dot{\phi}^2}~.
\end{equation}
From this, one can express the equation of state as
\begin{equation}
   w_\phi \equiv \frac{P_\phi}{\rho_\phi} = -1 \cdot \left[1-\dot{\phi}^2\right]~,
\end{equation}
which is constrained between $-1 < w_\phi < 0$. Due to the unusual form of the Lagrangian, the resulting equation of motion also takes a different form than the familiar Klein-Gordon equation. This equation of motion can be derived directly from the Lagrangian formalism or by requiring the usual comoving stress-energy conservation $\dot{\rho } + 3H (\rho+P) = 0$ with the Hubble parameter $H \equiv \dot{a}/a$ for a scale factor $a$. One finds
\begin{equation}\label{eq:eom}
    \frac{\ddot{\phi}}{1-\dot{\phi}^2}+3H\dot{\phi}+\frac{1}{V}\frac{dV}{d\phi}=0~.
\end{equation}
\enlargethispage*{2\baselineskip}
Note that $\dot{\phi} \to 1$ is dynamically prevented as the acceleration decreases the closer $\dot{\phi}$ gets to unity, and therefore we can safely assume $|\dot{\phi}| <1$ in all expressions. 
At this point it is interesting to look at several limiting cases. We observe that for $|\dot{\phi}| \ll 1$ the field behaves like a canonical scalar field (more on this in appendix \ref{app:canonical}).\footnote{Though it is required to re-normalize it as $\widetilde{\phi} = \int \sqrt{V} \mathrm{d}\phi$ in order to observe a truly canonical Lagrangian, stress-energy tensor, and equations of motion.} In particular, for $\dot{\phi} \to 0$, the field behaves like a cosmological constant and can therefore play the role of the dark energy. Interestingly, in the limit of $\dot{\phi} \to 1$, the energy density diverges while the pressure vanishes such that the field induces a dynamic equivalent to a dark matter dominated universe, a limit very different from that of a canonical scalar field. As such, a Tachyon field is a natural candidate to explain the late-time expansion history.

Importantly, throughout this work we are going to assume that the tachyon field is indeed responsible for the observed accelerated expansion of the universe, and remove any other dark energy/quintessence/cosmological constant from the cosmological model, resulting in what we will dub the $T$CDM model. The idea is that for sufficiently low kinetic velocities of the Tachyon, the potential can always be used to reach a given value of the observed Hubble constant today by rescaling the potential amplitude for a vanishing $\phi$, $V_0 = V(\phi=0)$.\footnote{For the polynomial potential, one instead rescales $V(\phi=1/N)$, see the corresponding section \cref{ssec:powerlaw}, and for the inverse exponential one instead rescales $V(\phi=1/\mu)$, see the corresponding section \cref{ssec:inverse_exp}.} More explicitly, we numerically adjust $V_0$ for each set of cosmological parameters such that the budget equation holds and \begin{equation}
    \Omega_T = \left[\frac{8\pi G}{3} \rho_\phi(z=0)\right] \cdot \frac{1}{H_0^2} \stackrel{!}{=}(1-\Omega_m - \Omega_r)~,
\end{equation}
with matter density fraction $\Omega_m$ and radiation density fraction $\Omega_r$\,. Investigations where the tachyon field only plays an observer field role are left for future work.

The interaction term of the DBI action for the tachyon field is naturally coupled to the electromagnetic sector, such that the fine-structure constant $\alpha$ becomes a function of the field itself as
\begin{equation}
    \alpha(\phi) = \frac{\beta^2 M_s^2}{2\pi V(\phi)}~,
\end{equation}
where $\beta$ is a warped factor, related to the conformation of the extra dimensions, and $M_s$ is the string mass scale \cite{Sen2002a,Sen2002b,Garousi2005, Copeland2005,Martins2016} (see also \cref{app:canonical}). As such, the relative variation of $\alpha$ becomes independent of $\beta$ and $M_s$ as
\begin{equation}
    \frac{\Delta \alpha}{\alpha_0} = \frac{V(\phi_0)}{V(\phi)}-1~,
\end{equation}
where $\phi_0$ is the tachyon field value today, and $\Delta \alpha = \alpha(\phi)-\alpha_0$ with $\alpha_0\approx 1/137$ being today's laboratory-measured value of the fine-structure constant. 

It is interesting to note that such an interaction arises naturally in this model without the need to explicitly postulate 
a given field-dependence for the coupling function (such as in the case of the $B_F(\phi)$ for dilaton-type models). As we are going to demonstrate below, this also means that the model is naturally restricted by cosmological and fine-structure constraints such that they prevent a fast evolution within the potential. In this limit, there is also a canonical field equivalent, as we detail in \cref{app:canonical}.

As a little side note, we conclude from \cref{eq:eom} (and already from \cref{eq:lagrangian}) that the units of the tachyon field are somewhat unusual. In particular, since $\dot{\phi}$ is dimensionless, we conclude that $\phi$ has units of time, or inverse energy (whereas a canonical scalar field is dimensionless, or in units of energy through the Planck mass), while the potential has the usual units of energy density. Therefore, the values of the field typically have to be viewed in relation to its coefficients within the potential, a point we make more explicit in the discussions below, see \cref{sec:models}.

All the freedom of this model lies in the choice of potential $V(\phi)$. In this work we are going to investigate a series of different potentials well motivated from an high energy perspective and assess their impact on cosmology. We will thus consider a set of $V(\phi)$ proposed in \cite{Garousi2005,Copeland2005} due to their cosmological relevance. Explicitly, we study all models of \cite{Copeland2005}. In the following sections, we list and discuss each of the potentials and their corresponding properties. In principle some of these potentials are dynamically unstable, see \cite{Garousi2005}, while their study remains relevant for this work.

\subsection{Data}

Our analysis of the tachyon models follows the numerical methodology already validated in previous works \cite{Vacher_2022,Vacher_2023,Schoneberg2023}. The tachyon equation of motion and its impact on the cosmological observables (including variations of the fine-structure constant $\alpha$) are implemented using a modified version of the \texttt{CLASS} Boltzmann-Solver \cite{Blas2011}.
We use the \texttt{MontePython v3.5} code for Bayesian MCMC inference \cite{Brinckmann2019} and produce plots using the \texttt{liquidcosmo} code.\footnote{Liquidcosmo is available at \url{https://github.com/schoeneberg/liquidcosmo}.} The adopted likelihoods include
\begin{itemize}
    \item \textbf{Planck:} CMB data from the 2018 (P-18) public data release (DR3), described further in \cite{Aghanim:2019ame}. Denoting temperature as T and E-mode polarization as E, we use the correlations of TT, TE, EE for high multipoles (high-$\ell$) using the \texttt{plik} likelihood, low-$\ell$ TT and EE data using the \texttt{commander}, as well as Planck lensing from \cite{Aghanim:2018oex}.
    \item \textbf{BAO:} We use the DR12 data from the SDSS-III BOSS survey \cite{Alam2017} as well as the SDSS main galaxy sample \cite{Ross:2014qpa} and the 6dFGS sample \cite{Beutler:2011hx}.\footnote{As these data do not play a pivotal role in setting the constraints, updating to newer releases such as the DESI BAO is unlikely to make a difference for our results.}
    \item \textbf{Supernovae:} We use all the 1701 un-calibrated supernova light curves (ranging from $z=0.001$ to $2.26$) from the Pantheon+ supernovae dataset \cite{Scolnic2022}.
    \item {\textbf{Cosmic chronometers:} We use the cosmic chronometers (measuring $H(z)$) from \cite{Moresco2022}, including the systematic covariance matrix of that work.}
    \item \textbf{Fine-structure data:} Data from the local universe constraining the fine-structure constant. This includes spectroscopic measurements from quasars \cite{Martins2017,alphaWebb,alphaSubaru,alphaespresso}, from the Oklo natural nuclear reactor \cite{Oklo}, and from laboratory measurements using atomic clocks from \cite{Filzinger2023}.
\end{itemize}

The full dataset will be hereafter named \enquote{All}, while the use of all the datasets without including fine-structure data will be dubbed \enquote{Cosmo}. While the constraints on the model parameters can be found in the respective sections, we also show the standard cosmological parameters for all cases in \cref{app:triangles} in \cref{fig:massive_cosmo,fig:invpow_cosmo,fig:anti_cosmo,fig:invexp_cosmo,fig:cosh_cosmo}. We note that all the chains have a Gelman-Rubin convergence criterion of $\left|R-1\right|<0.05$.

\section{Potentials under investigation and results}\label{sec:models}
\subsection{Massive rolling scalar potential}\label{ssec:massive}

\subsubsection{Description}
\begin{figure}
    \centering
    \includegraphics[width=\textwidth]{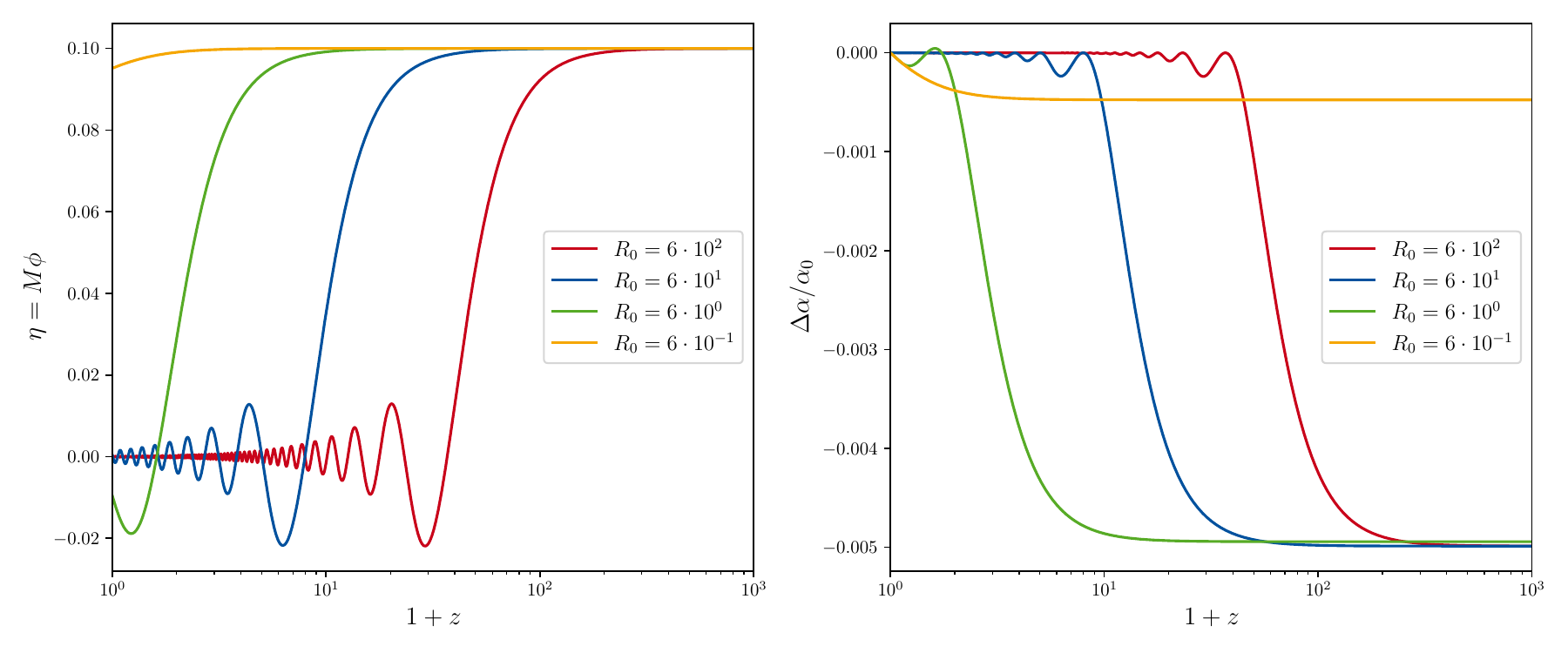}
    \caption{Massive rolling scalar potential field evolution and variation of the fine-structure constant. Note that both $\eta$ and $\Delta \alpha/\alpha_0$ almost perfectly scale with the initial value of $\eta$ ($\eta_\infty$), and therefore we fix $\eta_\infty = 0.1$.}
    \label{fig:massive_example}
\end{figure}

The massive rolling potential is given by
\begin{equation}
V\left(\phi\right)=V_0\exp\left[\frac{1}{2}M^2\phi^2\right]~.
\end{equation}
It represents an effectively massive potential (since $\ln V = \ln V_0 + \frac{1}{2} M^2 \phi^2$, and the equations of motion depend on $\mathrm{d} \ln V/\mathrm{d} \phi$). As such, we expect that the field begins rolling down towards the minimum at $\phi=0$ as soon as the Hubble-drag dilutes enough (when $H \sim \mathrm{d}\ln V/\mathrm{d}\phi$), which can in turn be translated to the condition $H \sim M$.

It is easy to see that the field $\phi$ can effectively be replaced by a field $\eta = M \phi$ that is dimensionless (though not canonical -- for the canonical limit of the rolling tachyon, see \cref{app:canonical}). In terms of this new variable the equation of motion \cref{eq:eom} becomes
\begin{equation}
    \frac{\eta''+\gamma \eta'}{R^2-(\eta')^2} + 3\eta' + \eta R^2  = 0~,
\end{equation}
with $R = M/H$ and $\gamma = \mathrm{d}\ln H/\mathrm{d} \ln a$, showing that the equations manifestly only depend on $R$, $\eta$ and $V_0$ (through \cref{eq:rho_P}) as well as their derivatives with respect to the logarithm of the scale factor, which we denote here by a $'$ symbol. Note that $|\dot{\phi}|<1$ enforces $|\eta'| < R$ (as can be seen also from the equations of motion). We can also write explicitly that
\begin{equation}\label{eq:delta_alpha_massive}
    \frac{\Delta \alpha}{\alpha_0} = \exp\left(\frac{\eta_0^2-\eta^2}{2}\right) - 1~,
\end{equation}
with $\eta_0 = \eta(z=0)$ and correspondingly $\alpha'/\alpha = - \eta \cdot \eta'$. Note that we can immediately deduce that fast moving fields are cosmologically excluded for this type of coupling, since to have $\Delta \alpha/\alpha \ll 1$ one needs $\eta' \ll 1$, whereas to have a fast moving field ($\eta' \to R$) one needs $R \gg \frac{3+\gamma}{2} \gtrsim \frac{3}{4}$, which is an obvious contradiction. More generally we expect fine-structure constant observations to tightly restrict the dynamics of this case.

In \cref{fig:massive_example} we show several evolutions for different values of $R_0 = M/H_0$ for a given value of the initial field $\eta_\infty$ in order to illustrate the typical dynamics. The point at which the field starts to become dynamical can be estimated using a very simplified model in which the tachyon late-time dynamics are ignored, leading to
\begin{equation}
    R \sim 1 \Rightarrow R_0 \sim \sqrt{\Omega_m (1+z)^3 + (1-\Omega_m)} \Rightarrow \ln(1+z) \sim \frac{2}{3} \ln R_0 - \frac{1}{3} \ln \Omega_m~,
\end{equation}
which fits very well the point at which the field has decreased to 80\% of its original value for a wide range of parameters.

\subsubsection{Constraints}

\begin{figure}
    \centering
    \includegraphics[width=0.49\textwidth]{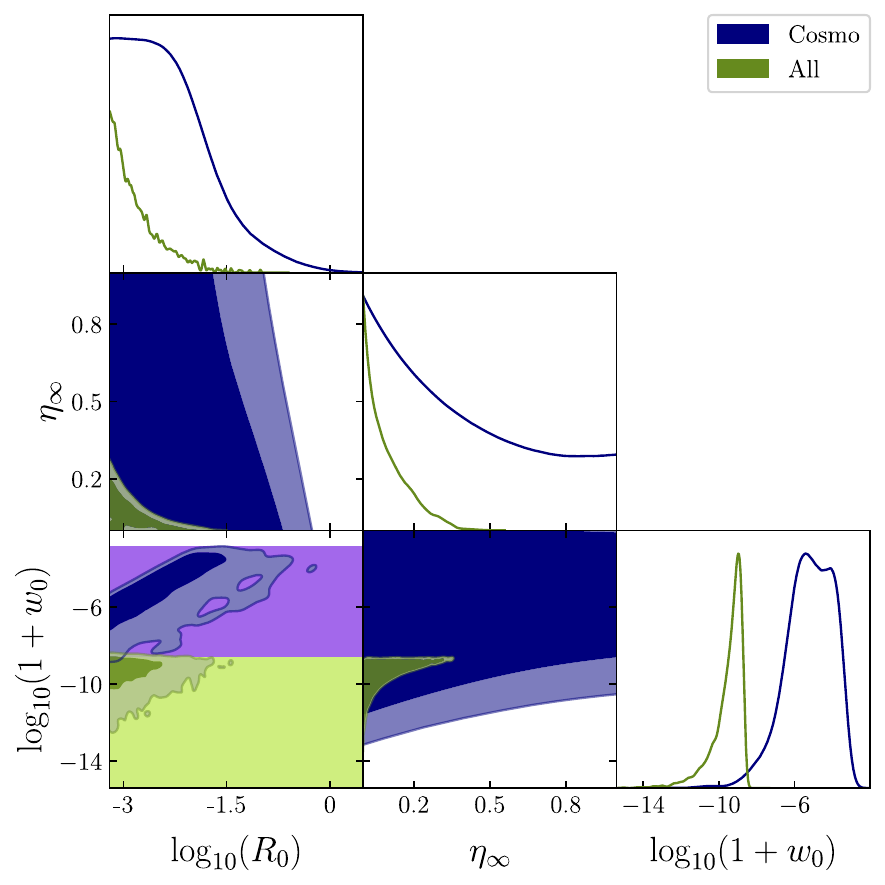}
    \includegraphics[width=0.49\textwidth]{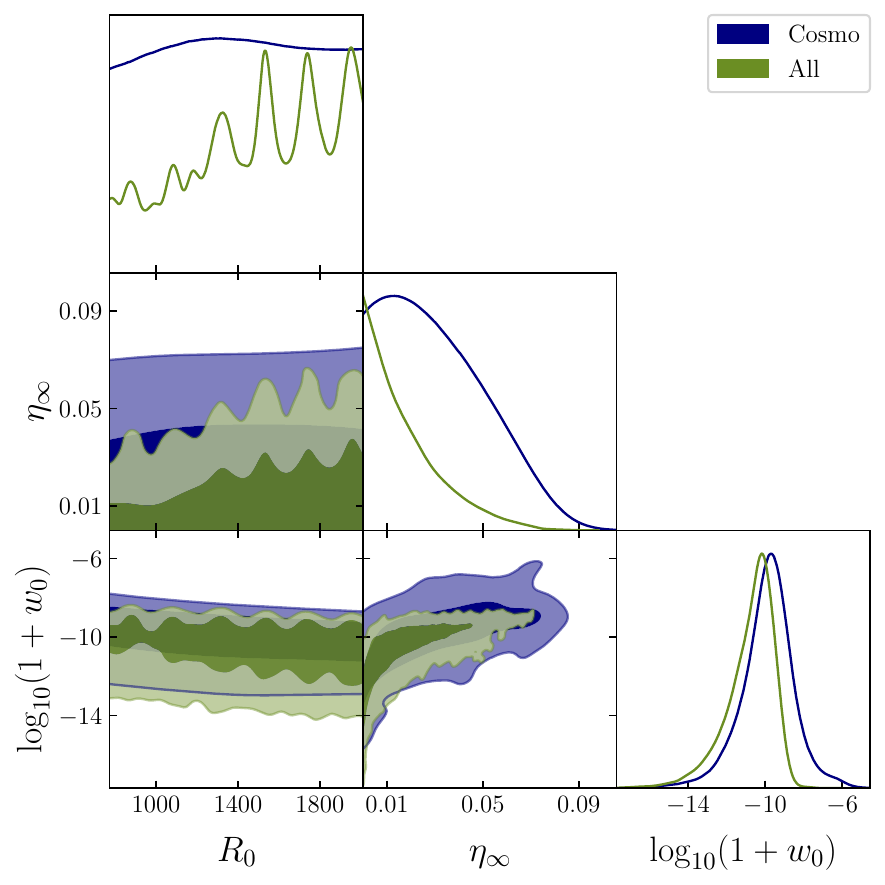}
    \caption{Constraints on the massive rolling scalar potential from cosmological and fine-structure data. We show the 68\% and 95\% marginalized constraints as dark and bright shaded regions. The blue color corresponds to cosmological data only, while the green color adds constraints from fine-structure constant data. The purple region in the bottom left panel shows where manual studies suggest that the contour would extend for larger priors on $\eta_\infty$ (see \cref{foot:purple}).}
    \label{fig:massive_alpha}
\end{figure}
The constraints on the parameter space of the massive rolling scalar potential can be found in \cref{fig:massive_alpha}. We show two regimes of the parameter $R_0$: one where $R_0 \lesssim 1$ corresponding to models with a mass so low that the Hubble drag remains comparatively large even until today, and one where $R_0 \gg 1$ for which models begin to oscillate around the minimum already at high redshift. In the former case the variation of the fine-structure constant is small because the field barely evolves and therefore the potential remains about the same, while in the latter case it is small because the field has evolved so much that it has moved to the bottom of the potential well, at which point the Hubble drag has damped the oscillations so much that the movement has stopped as well.

In the case of $R_0 \lesssim 1$ we find that for each value of $R_0$ there is an upper limit on $\eta_\infty$. This is due to higher values of $\eta_\infty$ resulting in larger field excursions and thus larger $\Delta \alpha/\alpha$. For the \enquote{All} case we see much tighter constraints than in the \enquote{Cosmo} case due to the very tight local fine-structure constraints. There is also a corresponding constraint on the value of $1+w_0$ (specifying the final velocity of the field). This constraint from the \enquote{Cosmo} case is at the level of around $10^{-3}$ compared to roughly $10^{-9}$ for the \enquote{All} data, a result that remains about the same for all potentials that cause late-time movement, as we show below.\footnote{There is a slight problem with very large $\eta_\infty$ that are difficult to numerically sample due to the squared exponential. We therefore added a purple region in the lower left hand panel of \cref{fig:massive_alpha} that highlights the $w_0$ values we would expect to be in principle accessible if higher $\eta_\infty$ values could be sampled.\label{foot:purple}}

Instead, for the case of $R_0 \gg 1$ we find that the constraints are mostly upper limits on $\eta_\infty$ given by the corresponding $\Delta \alpha/\alpha$ at CMB times or its variation today. Without fine-structure data the limit is caused by the $\Delta \alpha/\alpha$ at $z \gtrsim 1000$ that results from the variation between $V(\eta_\infty)$ to $V(\eta=0)$ and is therefore only mildly dependent on $R_0$. Similarly, with fine-structure data the upper limit is slightly tighter and caused by the tiny remaining field speed at $z \to 0$. This is also why the upper bound is very oscillatory in this regime, since if the oscillation happens to be at a minimum or maximum (giving a velocity close to zero) the fine-structure data bounds can be avoided, leaving only the cosmological limit. In principle each of the \enquote*{spikes} in the \enquote{All} data case would go up to the cosmological upper limit, but the MCMC only has a finite resolution for these spikes.

We suspect that even higher $R_0$ would cause the limit to gradually vanish since the oscillation would start before the CMB time and lead to the field already being sufficiently decayed to lead to no further oscillations (and a correspondingly vanishing field speed for the fine-structure data as well). However, the region is numerically difficult to explore due to the frequent rapid oscillations (and corresponding diverging runtime), which prevented further numerical explanation.

\subsection{Inverse power-law potential}\label{ssec:powerlaw}

\subsubsection{Description}

\begin{figure}
    \centering
    \includegraphics[width=\textwidth]{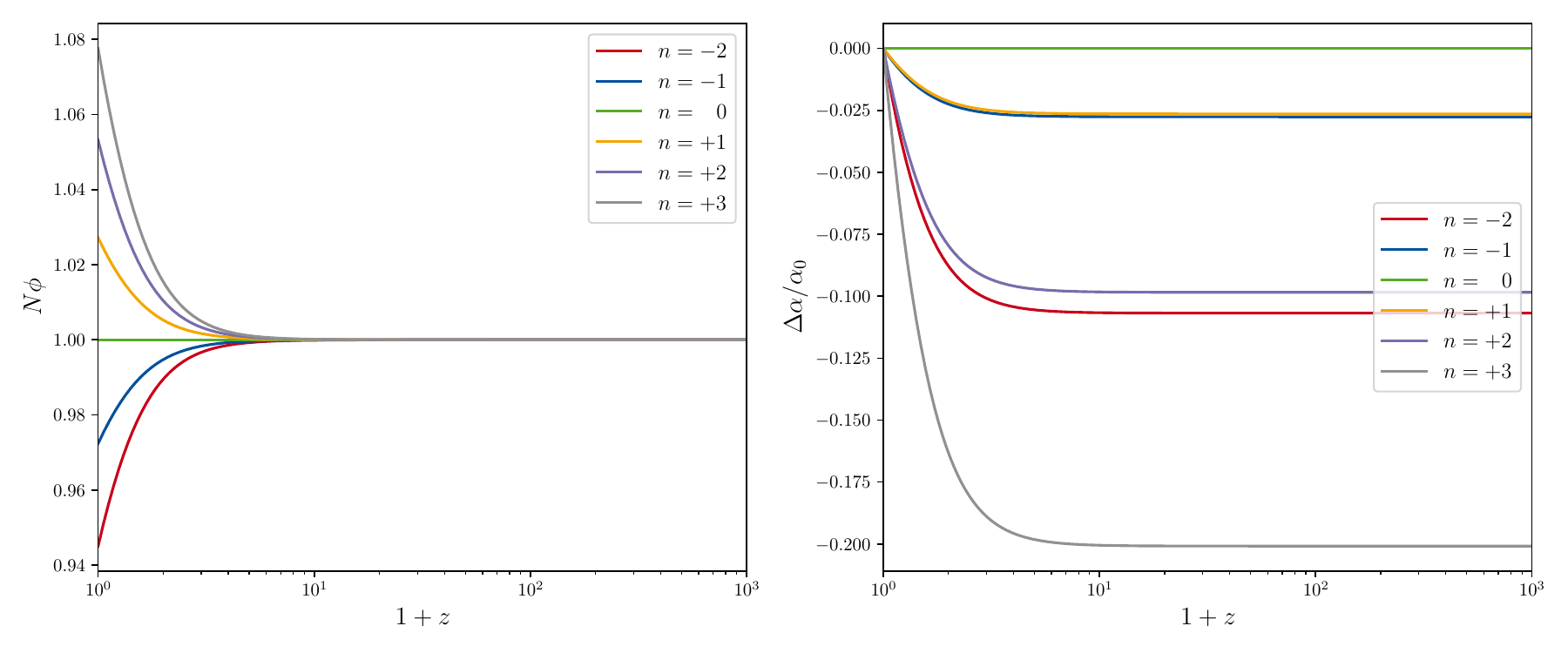}
    \caption{Example evolutions as in \cref{fig:massive_example} but for the inverse power-law potential. We set $\phi_\infty = 10^4\mathrm{Mpc} \approx 3.34 h/H_0$, balancing it with $N = 10^{-3}\mathrm{Mpc}^{-1}$ for easier visualization.}\label{fig:powerlaw_example}
\end{figure}
This potential is given by
\begin{equation}
    V(\phi)= V_0 (N\phi)^{-n}~,
\end{equation}
with an arbitrary constant $N$ and $V_0 = V(\phi=1/N)$ (since this case diverges as $\phi \to 0$). With this parameterization $N$ completely cancels out from the equations of motion and from the values of the fine-structure constant, and as such it does not have any physical significance. Its only purpose is to simplify the expressions. The equations of motion are
\begin{equation}
\frac{\ddot{\phi}}{1-\dot{\phi}^2} + 3 H \dot{\phi} - \frac{n}{\phi} = 0~. \label{eq:motion_powerlaw}
\end{equation}
Note that from this equation (since $\dot{\phi}$ must be unitless), it is clear that $\phi$ has the units of $1/H$, which for the chosen code will be Mpc. Since only ratios of $\phi$ are used for the expressions involving the fine-structure constant below, it is clear that such a choice of units has no observable consequences, though it should be kept in mind when interpreting the initial value ($\phi_\infty$).
For this case we have
\begin{equation}
    \frac{\Delta\alpha}{\alpha_0}=\left(\frac{\phi}{\phi_0}\right)^n-1~,
\end{equation}
and $\alpha'/\alpha = -n\phi'/\phi$. In \cite{Garousi2005,Copeland2005} it has been shown that this case provides an accelerated expansion asymptotically only as long as $n<2$, though in practice we find that even for values of $n\geq 2$ a transient acceleration can last long enough to be cosmologically interesting. Finally, for $n=0$ the model is completely equivalent to a cosmological constant and we expect no constraints on $\phi_\infty$ at that point. In \cref{fig:powerlaw_example} we show a few example evolutions of the field as a function of the $n$ parameter.
\begin{figure}
    \centering
    \includegraphics[width=0.6\textwidth]{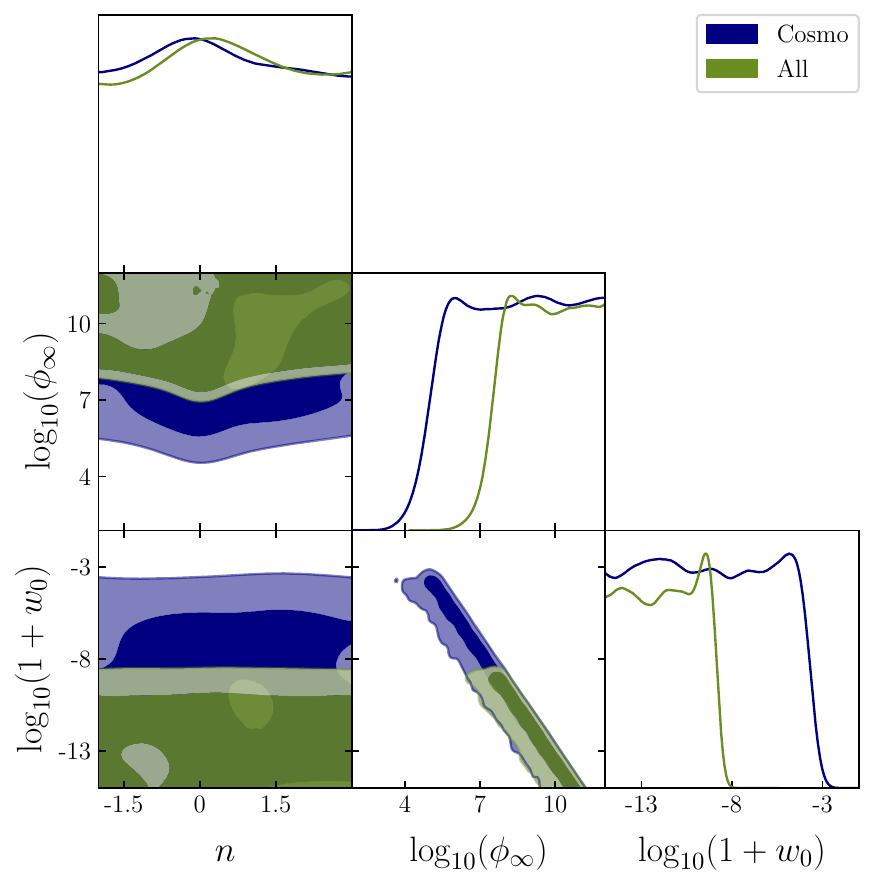}
    \caption{Constraints on the inverse power-law potential from cosmological and fine-structure data. We show the 68\% and 95\% marginalized constraints as dark and bright shaded regions. The blue color corresponds to cosmological data only, while the green color adds constraints from fine-structure constant data.}
    \label{fig:invpow_alpha}
\end{figure}
\subsubsection{Constraints}
As we observe in \cref{fig:invpow_alpha}, in this scenario we note that there is virtually no difference between the two datasets for the parameter $n$. The main difference is the lower limit on $\phi_\infty$ for each given value of $n$. For large values of $\phi_\infty$ the slope (according to \cref{eq:motion_powerlaw}) is tiny, allowing these solutions to have a vanishing motion independent of the precise value of $n$. We also observe that for $n \to 0$ the natural acceleration of the field is smaller, leading to the constraints on $\phi_\infty$ being weaker for $n \to 0$. Again, we expect a perfect logarithmic spike towards $n \to 0$, but this cannot be resolved with an MCMC algorithm. While the \enquote{Cosmo} case is dominated by the overall field movement causing a different $\alpha$ at CMB times compared to today, which the CMB data is sensitive to. Instead, the \enquote{All} case has the more restrictive fine-structure data putting constraints on the field velocity today, which is even more sensitive to the field evolution. Overall, there is also a tight relation between the $\phi_\infty$ value and the corresponding field velocity today (which can be quantified through $\log_{10}[1+w_0] = 2 \log_{10}[\dot{\phi}|_{z=0}]$), thus giving a tight constraint on $1+w_0$ as well, disallowing a strong departure from a behavior like a cosmological constant. As above, we find constraints from the \enquote{Cosmo} case at the level of around $10^{-3}$ compared to roughly $10^{-9}$ for the \enquote{All} case (we note that the apparent lower bound is simply an artifact of our prior choice and would appear for any finite upper value of $\phi_\infty$).

\subsection{Anti-massive potential}\label{ssec:antimassive}

\subsubsection{Description}
\begin{figure}
    \centering
    \includegraphics[width=\textwidth]{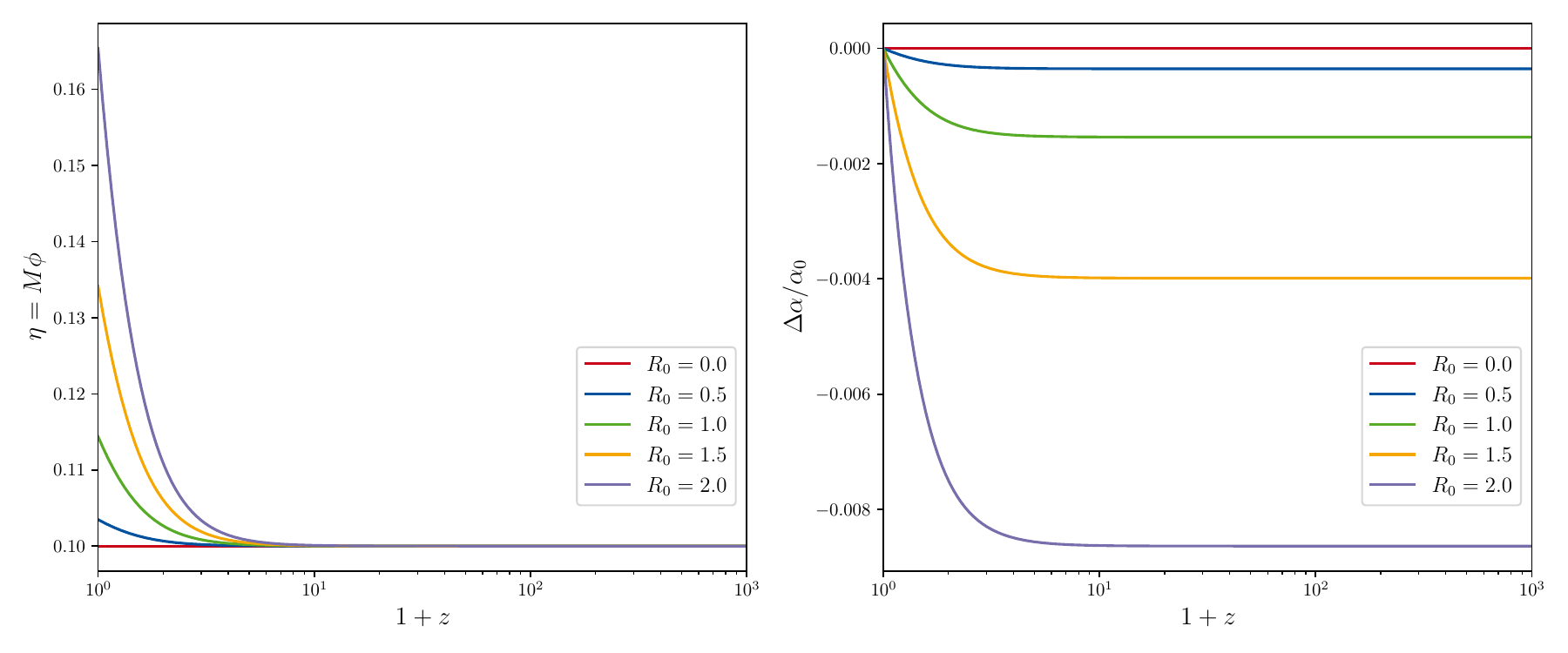}
    \caption{Example evolutions as in \cref{fig:massive_example} but for the anti-massive potential. We set $\eta_\infty = 0.1$ and test various values of $R_0$.}
    \label{fig:anti_example}
\end{figure}
The anti-massive potential is given by
\begin{equation}
    V\left(\phi\right)=V_0\exp\left[-\frac{M^2}{2}\phi^2\right]~,
\end{equation}
and can be seen as the direct equivalent to \cref{ssec:massive}. As such, we can employ the same transformations with $\eta = M \phi$ and defining $R = M/H$, $\gamma = \mathrm{d} \ln H/\mathrm{d} \ln a$. We have 
\begin{equation}
    \frac{\Delta\alpha}{\alpha_0} =\exp\left((\eta^2-\eta_0^2)\right)-1~,
\end{equation}
with $\alpha'/\alpha = \eta \cdot \eta'$. In \cref{fig:anti_example} we show a few example evolutions of the field as a function of the $R_0$ parameter. In this case the field naturally diverges from the origin at $\eta=0$ as soon as $R \sim 1$ during the evolution of the universe. Depending on how high $R_0$ is, this divergence becomes smaller or larger, but in any case is very likely to be extremely tightly constrained by local data due to this divergent behavior.

\subsubsection{Constraints}
\begin{figure}
\centering
    \includegraphics[width=0.6\textwidth]{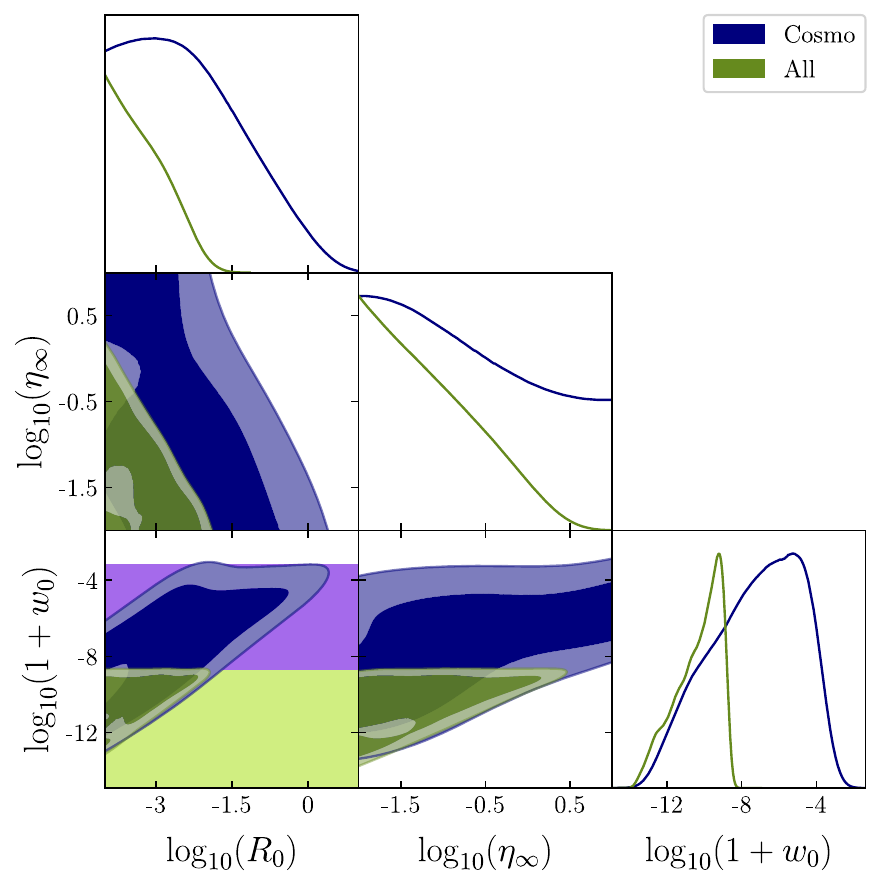}
    \caption{68\% and 95\% CL contours for the anti-massive potential case of \cref{ssec:antimassive}. In the 2D contours on the bottom row the purple/light-green regions show where manual studies suggest that the contour would extend for broader priors.\label{fig:anti_alpha}}
\end{figure}
The results for this case are displayed in \cref{fig:anti_alpha}. We empirically find that we can constrain $\eta_\infty \cdot R_0 < 10^{-1}$ with cosmological constraints, while we get $\eta_\infty \cdot R_0 < 10^{-4}$ for fine-structure constraints.

It is also important to note that the contours we see in \cref{fig:anti_alpha} are \enquote{incomplete} in the sense that we are restricted by the fact that we have finitely broad priors. In bottom left panel of \cref{fig:anti_alpha} the purple region shows a manually explored regime that we expect to be valid with broader priors, see also \cref{foot:purple}.\footnote{We note that simply broadening the priors is not numerically trivial here as the potential prefactor $V_0$ has to grow as $\exp(\tfrac{1}{2}\eta_\infty^2)$ to leave a order-unity potential initially, and in this regime one would require $\eta_\infty \simeq 10^2$, which gives $\exp(5\cdot 10^3)$, which is difficult to explore with double floating point precision.} In this case we also observe that cosmologically relevant deviations from a cosmological constant in terms of $1+w_0$ are excluded both by cosmological data (from the impact on the CMB) and from fine-structure data. As above, what we find from the \enquote{Cosmo} case is at the level of around $10^{-3}$ compared to roughly $10^{-9}$ for the \enquote{All} case. Also similarly to above, the lower bound on $1+w_0$ simply stems from choosing some lower bound on $\eta_\infty$ and is thus prior-driven.

\pagebreak[20]
\subsection{Negative exponential potential}\label{ssec:negative}

\subsubsection{Description}

\begin{figure}
    \centering
    \includegraphics[width=\textwidth]{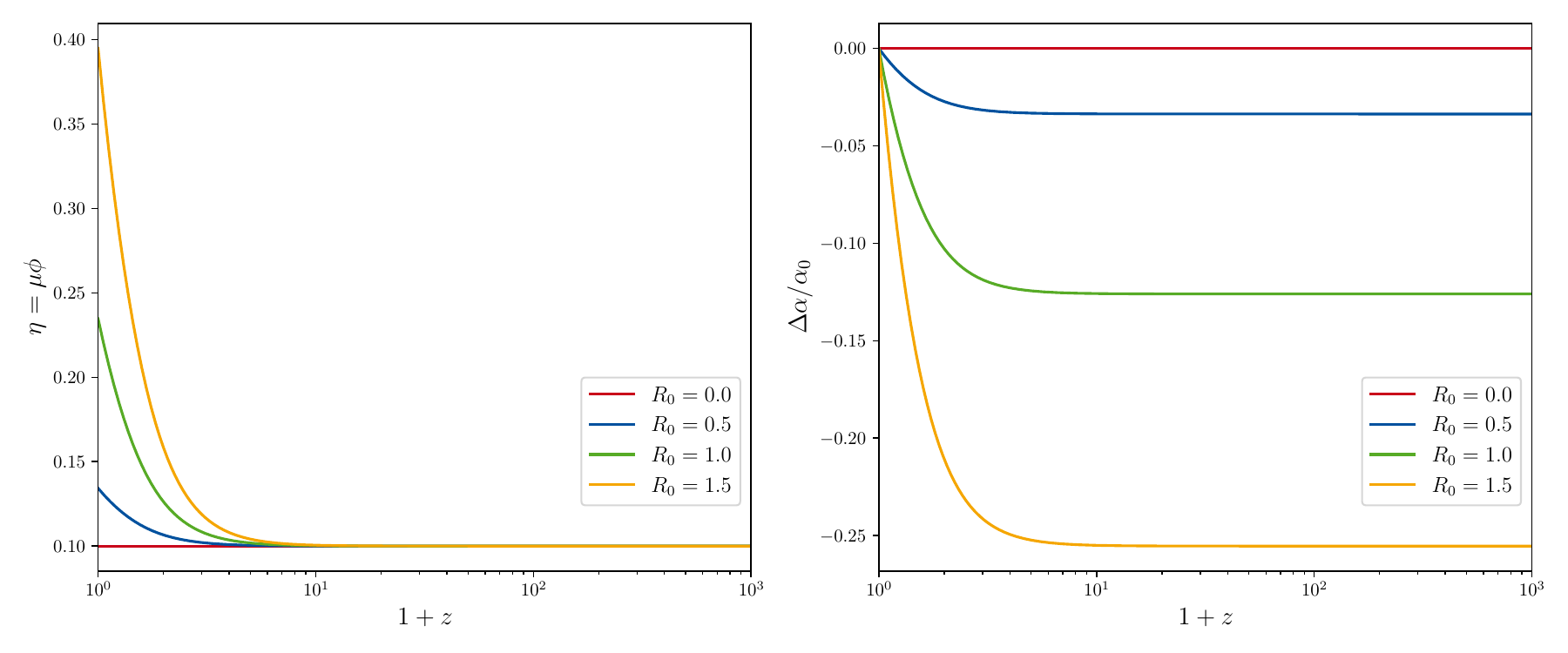}
    \caption{Example evolutions as in \cref{fig:massive_example} but for the negative exponential potential. We set $\eta_\infty = 0.1$ arbitrarily (see text about its irrelevance) and test various values of $R_0$.}
    \label{fig:negative_example}
\end{figure}
The negative expontential potential is given by
\begin{equation}
    V\left(\phi\right)=V_0\exp[-\mu\phi]~,\label{eq:potential_negative}
\end{equation}
leading us to introduce $\eta = \mu \phi$, which gives us the equations of motion
\begin{equation}
    \frac{\eta''+\gamma \eta'}{R^2-\eta'^2} + 3 \eta' - R^2 = 0~,
\end{equation}
with $R=\mu/H$ and $\gamma=\mathrm{d}\ln H/\mathrm{d}\ln a$ as defined before. Since this equation does not depend on $\eta$, it is actually a first order differential equation and can in principle be solved implicitly analytically, though the solution is not readily interpretable. Additionally, it means that the precise initial value of $\eta_\infty$ is irrelevant (and degenerate with $V_0$) for any observables. The variation of the fine-structure constant is simply
\begin{equation}
    \frac{\Delta \alpha}{\alpha_0} = \exp[\eta-\eta_0]-1~,
\end{equation}
and $\alpha'/\alpha = -\eta'$. In \cref{fig:negative_example} we show a few example evolutions of the field as a function of the $R_0$ parameter. In this case too the behavior is a divergence towards $\eta \to \infty$ for $\ln a \to \infty$, though within finite cosmological time the model can still be interesting.

\subsubsection{Constraints}

\begin{figure}
\centering
\includegraphics[width=0.5\textwidth]{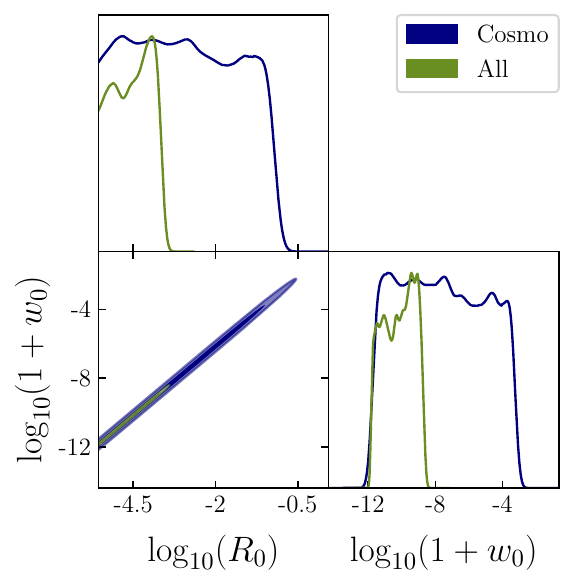}
\caption{68\% and 95\% CL contours for the negative exponential potential case of section \ref{ssec:negative}.}
    \label{fig:nega_alpha}
\end{figure}
We do not allow the initial field value in this case, since it is entirely degenerate with the potential amplitude, as evident from \cref{eq:potential_negative}. As such, we only allow $R_0$ to vary. Similarly to before we find an upper limit on $R_0$ from fine-structure data and a much broader limit on $R_0$ from cosmological data. This is shown in \cref{fig:nega_alpha}. As expected, cosmological data exclude order unity $R_0$ leading to $1+w_0 \sim 10^{-3}$, while fine-structure data constraints are much tighter (in this case $10^{-3}$), leading to stronger constraints of $1+w_0 \sim 10^{-9}$.

\begin{figure}
    \centering
    \includegraphics[width=\textwidth]{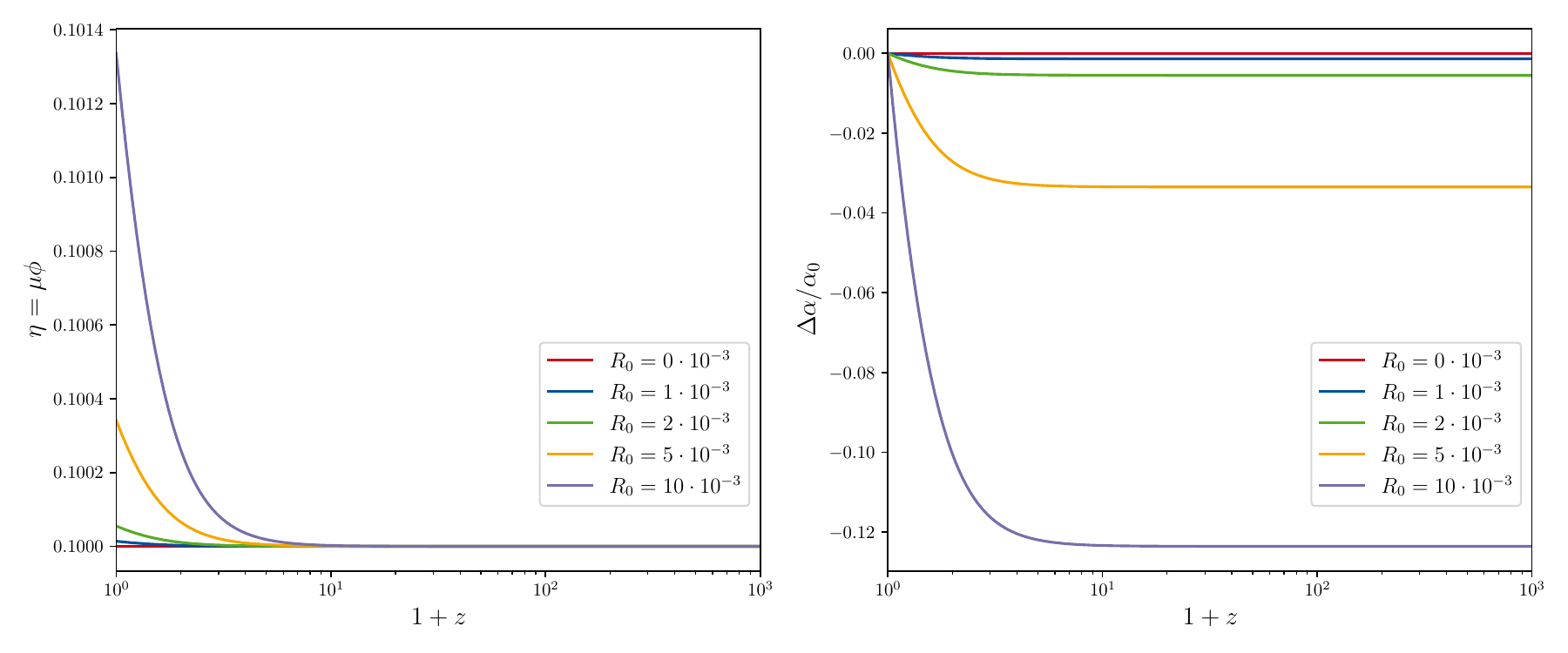}
    \caption{Example evolutions as in \cref{fig:massive_example}, but for the inverse exponential potential. We set $\eta_\infty = 0.1$ and test various values of $R_0$.}
    \label{fig:invexp_example}
\end{figure}
\clearpage
\subsection{Inverse exponential potential}\label{ssec:inverse_exp}

\subsubsection{Description}

The inverse potential is given by
\begin{equation}
    V\left(\phi\right)=V_0\exp[1/(\mu\phi)]~.
\end{equation}
Defining once again $\eta =\mu \phi$ and $R = \mu/H$, we arrive at the equations of motion as
\begin{equation}
    \frac{\eta''+\gamma \eta'}{R^2-\eta'^2} + 3 \eta' - \frac{R^2}{\eta^2} = 0~,
\end{equation}
and the variation of the fine-structure constant
\begin{equation}
    \frac{\Delta\alpha}{\alpha_0}=\exp\left[\frac{1}{\eta_0} - \frac{1}{\eta}\right]-1~,
\end{equation}
with $\alpha'/\alpha = \eta'/\eta^2$. In \cref{fig:invexp_example} we show a few example evolutions of the field as a function of the $R_0$ parameter.  Here too the field naturally runs to infinity as $\ln a \to \infty$ and the cosmological interest is derived from the finite evolution.

\subsubsection{Constraints}

\begin{figure}
\centering
\includegraphics[width=0.49\textwidth]{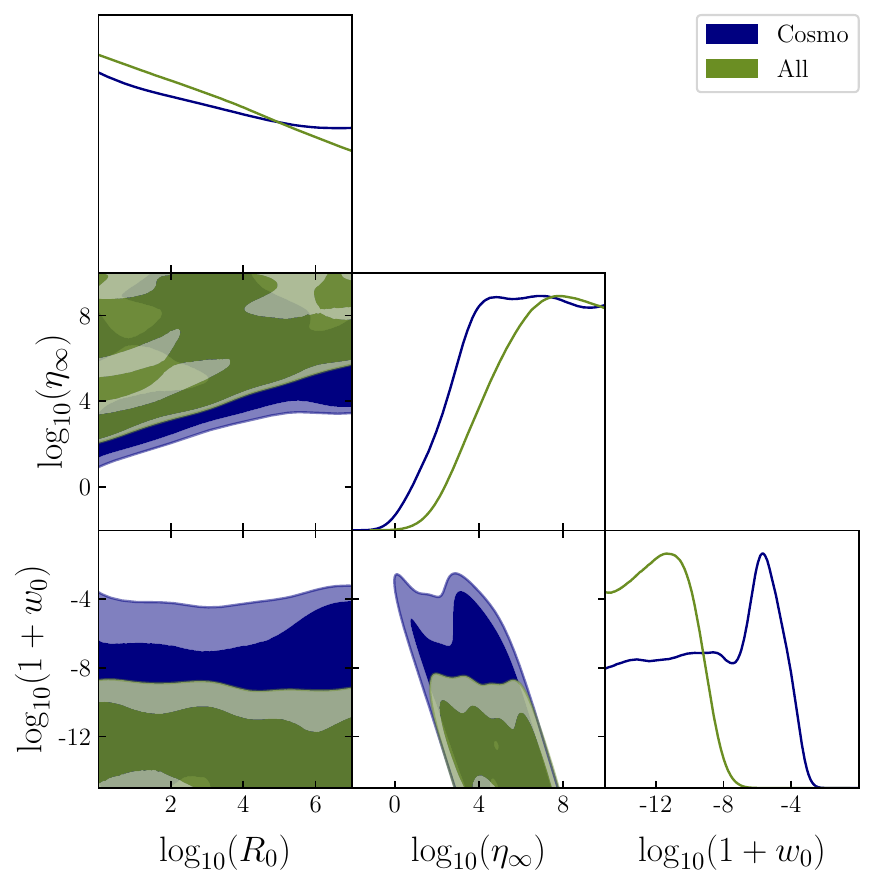}
\caption{Same as \cref{fig:invpow_alpha}, but for the inverse exponential potential instead.}
    \label{fig:invexp_alpha}
\end{figure}
In this scenario we see the largest disparity between the two datasets. We can see in \cref{fig:invexp_alpha} that fine-structure data imposes very strong constraints. For each value of $R_0$ we find a lower bound on $\eta_\infty$, which is much tighter when considering fine-structure data. This has to do with the extremely sharp dependence on $\eta_\infty$ in the equations of motion. The bounds are approximately given by $\log_{10}(R_0/\eta_\infty^2) < 1.4$ for cosmological data and $\log_{10}(R_0/\eta_\infty^2) < -3.2$ for fine-structure data. Additionally, we find approximately the same fine-structure and cosmological bounds as above, namely $1+w_0 < 10^{-9}$ from fine-structure constraints and $1+w_0 < 10^{-3}$ from cosmological constraints.

\subsection{Cosh potential}

\subsubsection{Description}

\begin{figure}
    \centering
    \includegraphics[width=\textwidth]{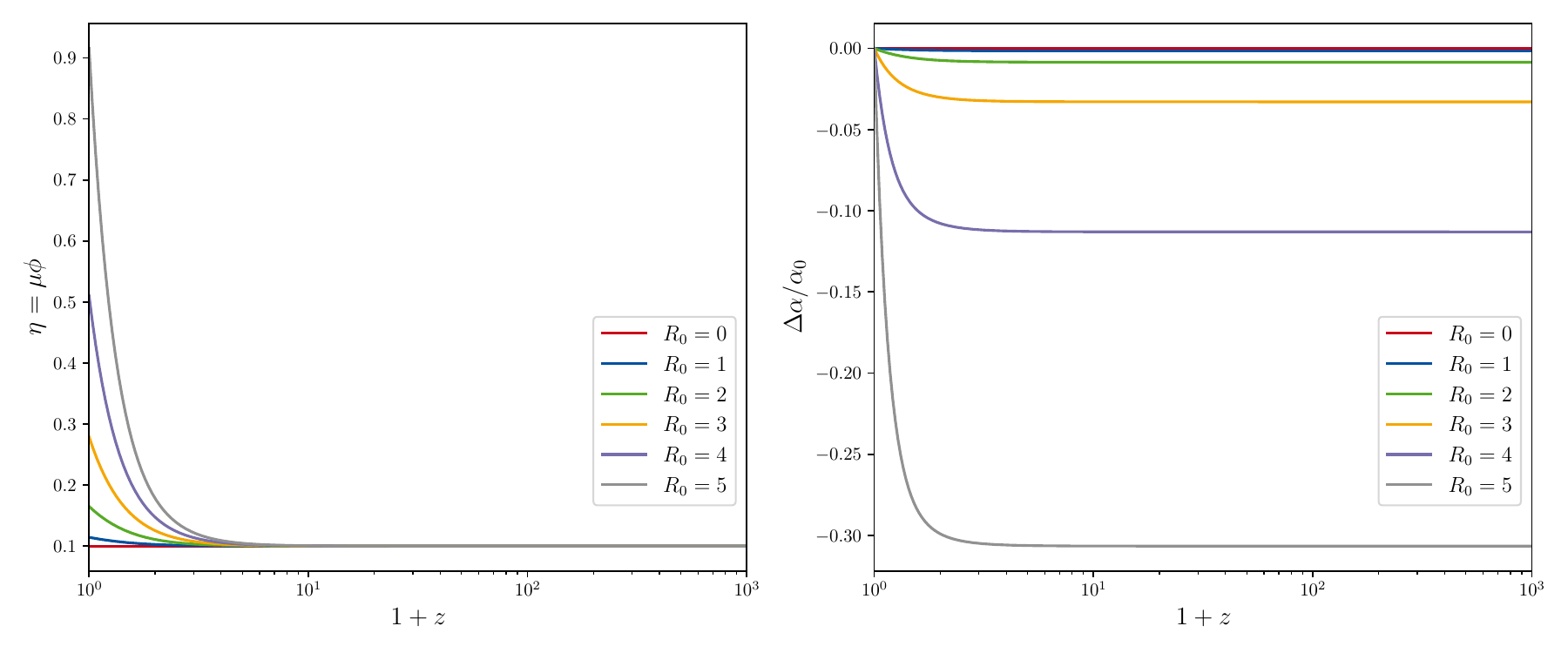}
    \caption{Example evolutions as in \cref{fig:massive_example} but for the cosh potential. We set $\eta_\infty = 0.1$ and test various values of $R_0$. We see that the $\Delta \alpha/\alpha_0$ diverges exponentially with larger values of $R_0$.}
    \label{fig:cosh_example}
\end{figure}
The cosh potential is given by
\begin{equation}
    V\left(\phi\right)=\frac{V_0}{\cosh\left(\mu\phi\right)}~.
\end{equation}
Here we also define $\eta = \mu \phi$. In terms of this variable the equations of motion become
\begin{equation}
    \frac{\eta''+\gamma \eta'}{R^2-\eta'^2} + 3 \eta' - R^2 \tanh(\eta) = 0~,
\end{equation}
where $R = \mu/H$ as above. The variation of the fine-structure constant is 
\begin{equation}
    \frac{\Delta \alpha}{\alpha_0} = \frac{\cosh(\eta)}{\cosh(\eta_0)}-1~,
\end{equation}
and $\alpha'/\alpha = \eta' \tanh(\eta)$. In \cref{fig:cosh_example} we show a few example evolutions of the field as a function of the $R_0$ parameter. These show the same kind of strong divergence of the field towards $z \to 0$, which in this case strongly diverges as $R_0 > 1$.

\subsubsection{Constraints}

\begin{figure}
    \centering
    \includegraphics[width=0.49\textwidth]{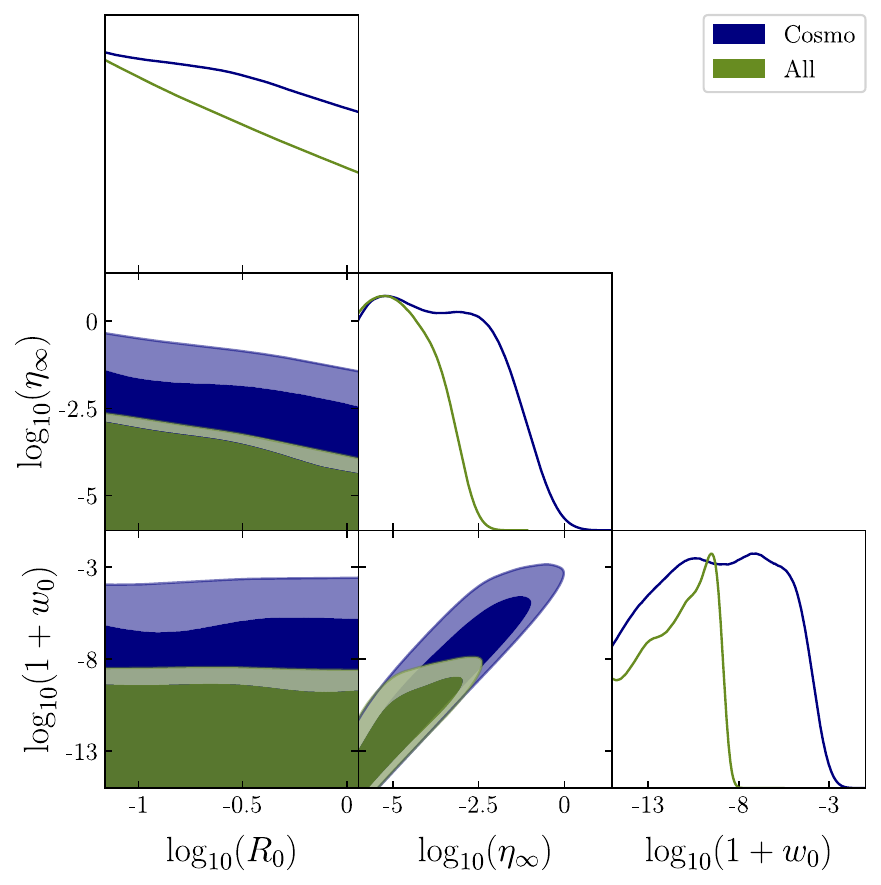}
    \includegraphics[width=0.49\textwidth]{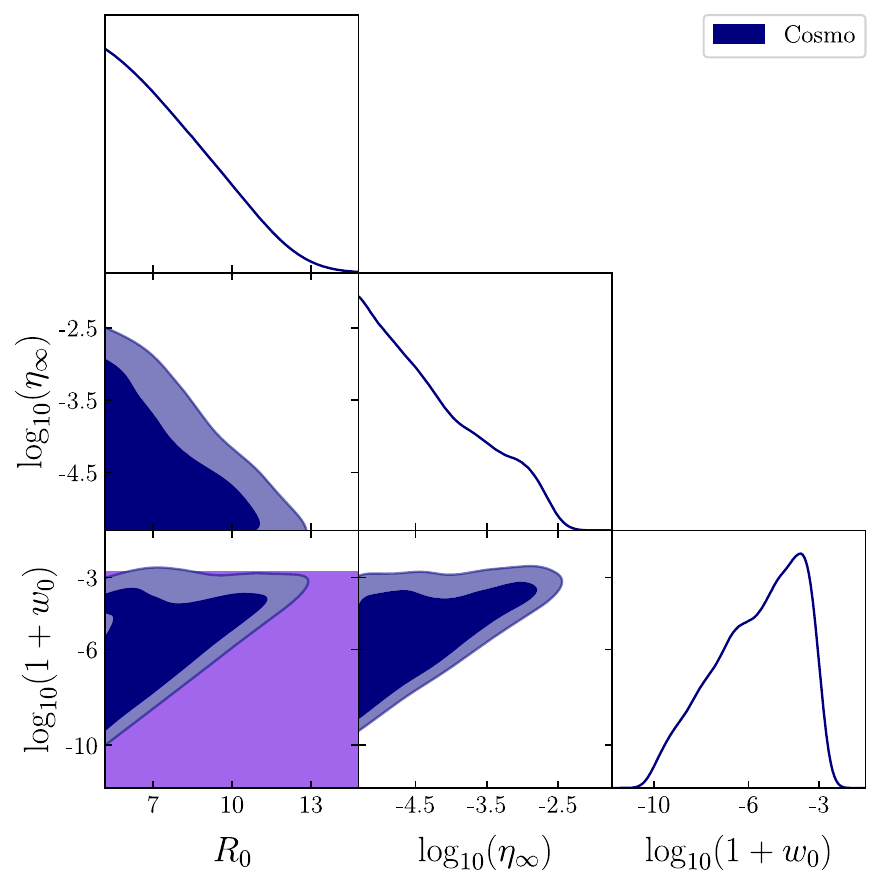}
    \caption{Same as \cref{fig:massive_alpha}, but for the cosh potential instead. The right panel shows the high $R_0$ limit, in which the cosmological constraints begin to decay exponentially as a function of $R_0$.}
    \label{fig:cosh_alpha}
\end{figure}
Now we present the constraints for the final scenario in this study. As shown in  the left panel of \cref{fig:cosh_alpha}, we observe that we find an upper limit on $\eta_\infty$ for each value of $R_0$. Interestingly, neither in the cosmological data case nor in the fine-structure constant case we find order unity values of $R_0$ excluded. Despite this, we do find the same $w_0 \sim 10^{-3}$ and $w_0 \sim 10^{-9}$ bounds as above for the \enquote{Cosmo} and \enquote{All} cases, respectively. The reason that larger values of $R_0$ are allowed for this particular potential compared to the negative and the massive case is somewhat surprising, and the origin of this relaxed bound is not quite clear, given the vast similarity of the equations of motion.

In any case, we investigate also the regime of higher $R_0$ values for this reason, and the results are shown in the right panel of \cref{fig:cosh_alpha}. There we find that at large $R_0$ there is an exponential dependence of the initial field offset with respect to the value of $R_0$. We find that approximately $2 \log_{10}(\eta_\infty) + R_0 < 2.6$, and we further confirm the cosmological bound at $1+w_0 < 10^{-3}$ as expected.

\section{Discussion}\label{sec:discussion}

\begin{table}
    \centering
    \resizebox{\textwidth}{!}{%
    \begin{tabular}{c c|c c|c c}
         & & \multicolumn{2}{c|}{\enquote{Cosmo}} & \multicolumn{2}{c}{\enquote{All}} \\
        Potentials & Constraint type & \rule{0pt}{2.5ex} Constraint & $\log_{10}(1+w_0)$ & Constraint & $\log_{10}(1+w_0)$\\ \hline\rule{0pt}{2.8ex}
        \rule{0pt}{2.8ex}Massive (left)& $\log_{10}(R_0)+\log_{10}(\eta_\infty)$& $<-0.84$ & $<-2.8$ & $<-3.7$ & $<-8.5$ \\ \rule{0pt}{2.8ex}
        \rule{0pt}{2.8ex}Massive (right)& $\eta_\infty$& $<+0.087$ & $<-6.6$ & $<+0.058$ & $<-8.6$ \\ \rule{0pt}{2.8ex}
        Inverse power-law & $\log_{10}(|n|)-\log_{10}(\phi_\infty)$& $<-4.5$ & $<-2.9$  & $ <-7.2$ & $<-8.3$ \\ \rule{0pt}{2.8ex}
        Anti-massive &$\log_{10}(R_0)+\log_{10}(\eta_\infty)$& $<-0.77$ & $<-2.7$ & $<-3.6$ & $<-8.4$ \\ \rule{0pt}{2.8ex}
        Negative & $\log_{10}(R_0)$&$ <-0.83$ & $<-2.9$  & $ <-3.7$ & $<-8.5$  \\ \rule{0pt}{2.8ex}
        Inverse exponential &$\log_{10}(R_0)-2\log_{10}(\eta_\infty)$ & $<+1.6 $ & $<-2.9$ & $<-3.2 $ & $<-7.6$ \\ \rule{0pt}{2.8ex}
        Cosh (left)&$\log_{10}(R_0)-\log_{10}(\eta_\infty)$& $<-0.40$ & $<-2.8$ & $<-3.0$ & $<-8.3$ \\ \rule{0pt}{2.8ex}
        \rule{0pt}{2.8ex}
        Cosh (right)&$R_0+2 \log_{10}(\eta_\infty)$& $<+2.6$ & $<-2.4$ & \noindent\rule{15pt}{0.1ex} & \rule{15pt}{0.1ex}
    \end{tabular}
    }
    \caption{Upper bounds on $1+w_0$ and for the parameters of the potential for each investigated potential. In the first column we state the potential, in the second column which kind of constraint we report on in the third and fifth column. These correspond to the physical constraints imposed by the data for combinations of the underlying parameters. Instead of giving the usual $95\%$ CL upper bound, we instead quote where the posterior has dropped to below $5\%$ of its maximum value. This way of quoting the upper limit is more resistant to the arbitrary choice of priors on the underlying parameters.}
    \label{tab:constraints}
\end{table}

\begin{figure}
    \centering
    \includegraphics[width=0.49\textwidth]{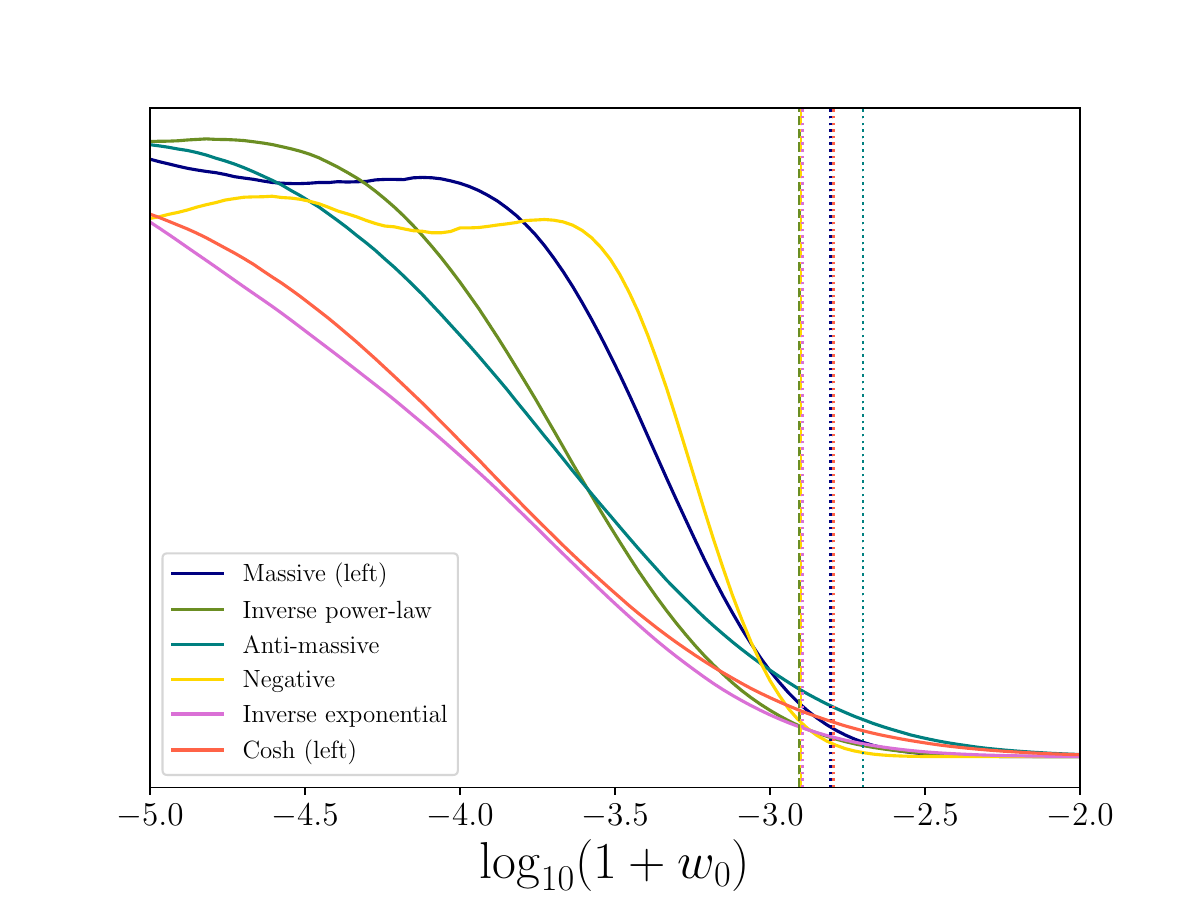}
    \includegraphics[width=0.49\textwidth]{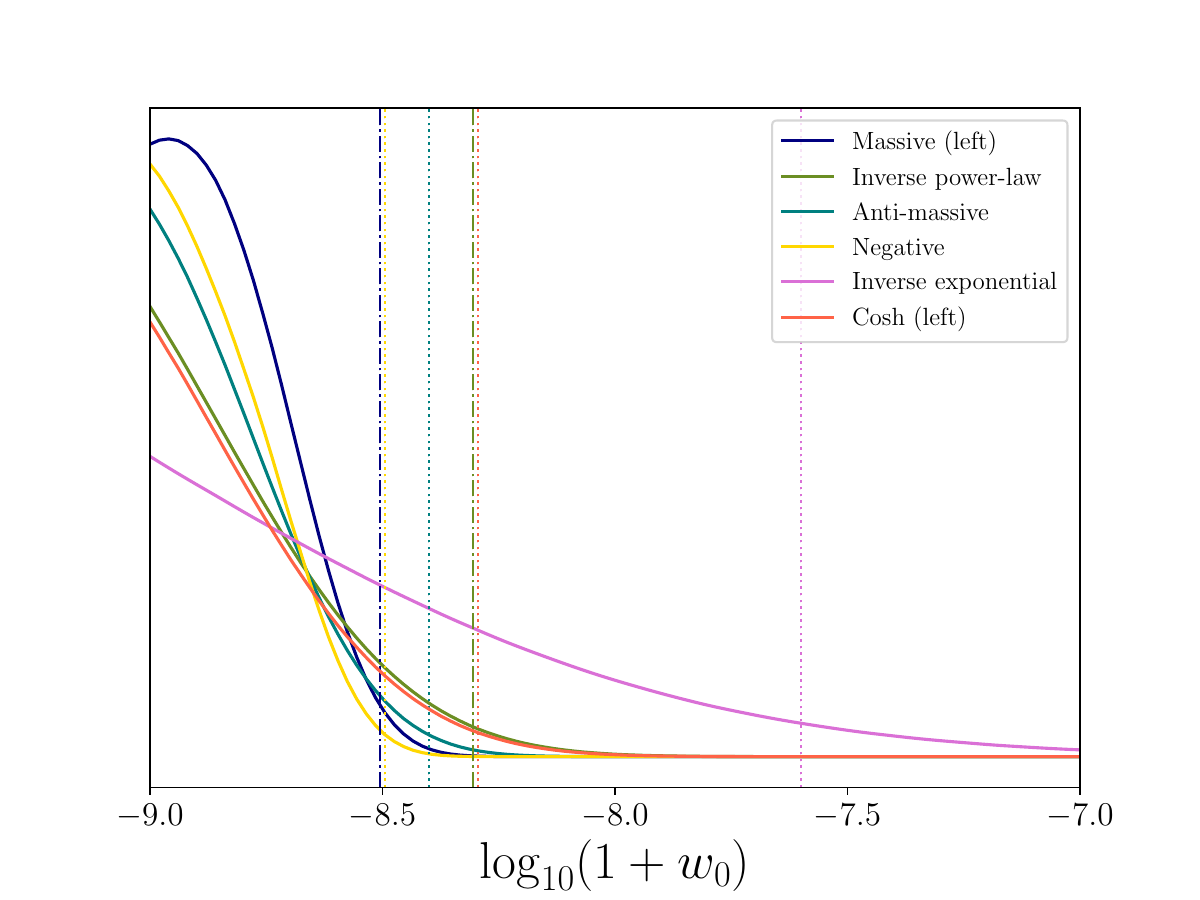}
    \caption{Posterior distribution for all the potentials in study (named according to \cref{tab:constraints}). The left plot represents the \enquote{Cosmo} dataset and the right one the \enquote{All} dataset. The dashed lines represent the upper limits on \mbox{$\log_{10}(1+w_0)$} for each potential as reported in \cref{tab:constraints}.}
    \label{fig:all}
\end{figure}

While the rolling tachyon has been motivated for a wide variety of potentials, in this work we demonstrate that there are several general conclusions that can be drawn for this type of solution. First, as summarized in \cref{tab:constraints} and \cref{fig:all}, we find that the equation of state today $w_0 = w_\phi(z=0)$ is generically constrained in these potentials. Below, we show a simple analytical exploration of the tachyon field that allows us to build an intuition about the general behavior and the expected constraints.

In the limit of $\dot{\phi} \ll 1$ (such as at initial time), we can expand the equations of motion (from \cref{eq:eom}) using
\begin{equation}\label{eq:eom_approx}
    \frac{\mathrm{d}\dot{\phi}}{\mathrm{d}\ln a} + 3 \dot{\phi} + \frac{\mathrm{d}\ln V/\mathrm{d}\phi}{H} =0~,
\end{equation}
which permits a solution with $\dot{\phi} \ll 1$ as long $|\mathrm{d}\ln V/\mathrm{d}\phi| \ll H$ (which is typically fulfilled in the early universe). As such, during the initial time the field is typically frozen until $H \sim |\mathrm{d}\ln V/\mathrm{d}\phi|$, at which point it begins to accelerate down the potential (one can use the stress-energy conservation equation with $w_\phi>-1$ to show that $\dot{\rho_\phi}<0$ and, as long as $\dot{\phi} \ll 1$, this implies that $\mathrm{d}V(\phi)/\mathrm{d}t < 0$). Therefore, initially $\alpha$ must increase,\footnote{Since the field rolls typically to lower regions of the potential, one generically expects $V(\phi_0)<V(\phi)$, and therefore $\Delta \alpha/\alpha<0$ throughout, not just initially. This is what we observe, and was also discussed in \cite{Martins2016}.} as evident in all of our cases, see \cref{fig:massive_example,fig:anti_example,fig:negative_example,fig:powerlaw_example,fig:invexp_example,fig:cosh_example}. This increase in $\alpha$ continues unavoidably unless the field can roll into a plateau or minimum of the potential. With the additional condition that the field make up the dark energy responsible for the accelerated expansion of the universe, this plateau or minimum is not allowed to be at $V(\phi) \to 0$. The only case obeying this condition is the one with the massive potential, which is indeed the only model which permits a non-trivial parameter space (corresponding to $R_0 \gg 1$, see \cref{ssec:massive}). Without this condition, the increase of $\alpha$ must have only begun close to today, as otherwise $\alpha_\mathrm{cmb}/\alpha_0$ would become too large. This is exactly what happens in \cref{fig:anti_example,fig:powerlaw_example,fig:negative_example,fig:invexp_example,fig:cosh_example}.

To illustrate cases of this kind further, we can explicitly approximate the local patch of the potential where the field begins to accelerate. At first order we can expand 
\begin{equation}
     \ln V(\phi) = \ln V_0 + \kappa (\phi-\phi_\infty) + \mathcal{O}([\phi-\phi_\infty]^2)~.
\end{equation}
This approximation is based on $\dot{\phi} \ll 1$ initially, allowing for such an expansion. In this local regime the approximate equations of motion of \cref{eq:eom_approx} imply
\begin{equation}
    \frac{\mathrm{d}\dot{\phi}}{\mathrm{d}\ln a} + 3 \dot{\phi} + \frac{\kappa}{H} \approx 0~,
\end{equation}
The solution involves an irrelevant decaying term ($\propto a^{-3}$) and a growing term
\begin{equation}
    \dot{\phi} \approx - \frac{\kappa}{H_0} \int_{-\infty}^{\ln a} \frac{\tilde{a}^3}{E(\tilde{a})} \mathrm{d}\ln \tilde{a}~,
\end{equation}
with the expansion rate $E(a) = H(a)/H_0$\,. In particular, we find that $\dot{\phi} \sim \kappa/H_0$ \,. We explicitly find that $1+w_0 = (\dot{\phi}|_{z=0})^2 \sim (\kappa/H_0)^2$
\begin{equation}
    \frac{\mathrm{d}\ln \alpha}{\mathrm{d}\ln a} = - \frac{\mathrm{d}\ln V}{\mathrm{d} \ln a} \approx - \frac{\kappa}{H} \dot{\phi} \sim \kappa^2/H_0^2~,
\end{equation}
and
\begin{equation}
\begin{aligned}
    \Delta \alpha/\alpha_0 &= \frac{V(\phi_0)}{V(\phi)}-1 \approx \exp[-\kappa (\phi-\phi_0)]-1 \approx -\kappa (\phi-\phi_0) = -\kappa \int_t^{t_0} \dot{\phi} \mathrm{d}t  \\& = -\kappa \int_{-\infty}^{\ln a} \dot{\phi}/H(\tilde{a}) \mathrm{d} \ln \tilde{a} \sim \kappa^2/H_0^2~.
\end{aligned}
\end{equation}
We recognize that under this approximation there is a very tight link between $1+w_0$, $\mathrm{d}\ln \alpha/\mathrm{d} \ln a$, and $\Delta \alpha/\alpha_0$. Indeed, the analytical argument tells us that they should be all of the same order of magnitude. This is indeed what we find. The CMB can constrain $\Delta \alpha/\alpha_0 \sim 10^{-3}$ \cite{Aghanim:2019ame}, corresponding to the \enquote{Cosmo} constraints of $1+w_0 \sim 10^{-3}$, whereas fine-structure data constrain $\mathrm{d}\ln \alpha/\mathrm{d} \ln a \sim 10^{-9}$ leading to \enquote{All} constraints of $1+w_0 \sim 10^{-9}$. Since such divergence in this type of potential is not bounded, the motion necessarily needs to happen towards $z \to 0$ as otherwise the constraints would be further violated.

To summarize, we find that solutions with divergent $\alpha$ can only be avoided whenever the potential has a minimum or plateau which is not at $V(\phi)\to 0$. If this is not the case, we find that we expect to have $1+w_0$, $\mathrm{d}\ln \alpha/\mathrm{d} \ln a$, and $\Delta \alpha/\alpha_0$ all of the same order of magnitude, leading to the natural expectation that CMB data will give constraints on $1+w_0 \sim 10^{-3}$ and fine-structure data $1+w_0 \sim 10^{-9}$, restricting the Tachyon field to not have a significant departure from a behavior as a cosmological constant.

As we further discuss in \cref{app:eotvos}, the additional use of free-fall data such as the latest result of the MICROSCOPE mission \cite{Touboul2022} might even further narrow the bounds on the deviation from a cosmological constant down to $1+w_0 \sim 10^{-11}$. Including these constraints explicitly (going beyond an order-of-magnitude estimate) turns out to be highly non-trivial, since they would require a generalization of the derivation of the E\"otv\"os parameter from canonical massive fields to tachyon fields with arbitrary potentials. While the field is constrained to have a canonical evolution (due to $\dot{\phi} \ll 1$ as we show above, see also \cref{app:canonical}) the generalization of the fifth-force interactions for a non-massive Klein-Gordon equation is left for future work. However, the rolling tachyon model is now constrained to be so indistinguishably close to the $\Lambda$CDM by $\alpha$ data that such an improvement of the result would not have changed our conclusions.

We also mention in passing that the directions of the tightest constraint (shown in \cref{tab:constraints}) for most cases correspond to those that are effectively equivalent to constraints on $|\mathrm{d}\ln V/\mathrm{d}\phi/H|$ at $z\to0$, which according to the above discussion is roughly equivalent to constraints on $1+w_0$\,, indirectly confirming our argument.

\section{Conclusion}\label{sec:concl}

Rolling tachyon fields are a type of non-canonical scalar field arising in string theory with a natural variations of the fine-structure constant. In this work we have investigated a variety of proposed potentials for this type of scalar field, and we find that we can obtain generic constraints on these kinds of models. Using cosmological datasets only (CMB, BAO, supernovae) we find constraints for the deviation of a cosmological constant (specified as $1+w_0 = 1+[P_\phi/\rho_\phi]|_{z=0} = [\dot{\phi}|_{z=0}]^2$) of the order $1+w_0 \sim 10^{-3}$ and using fine-structure data we find $1+w_0 \sim 10^{-9}$. We have demonstrated that these constraints naturally arise in any potential that does not have a minimum or plateau at $V(\phi)>0$. In terms of the model parameters, this typically results in $R_0 \sim \mathcal{O}(1)$, though this depends on the precise shape of the potential.

We also note that such models have $\dot{\phi} \ll 1$ and therefore are mostly equivalent to canonical scalar fields with a special coupling. As such, we also find a connection to the arguments of \cite{Vacher2024}, which states that canonical scalar fields cannot produce a cosmologically relevant variation of the fine-structure constant at the epoch of recombination due to the strong constraints from fine-structure data. We have now verified explicitly that despite their non-canonical couplings and evolution, the rolling tachyon models obey the same arguments.

The massive potential is the only example of the potentials proposed in \cite{Copeland2005} that can avoid such constraints, and we find a non-trivial part of the parameter space. However, here too we find $1+w_0 < 10^{-8}$ due to the ring-down of the scalar field evolution at late times, see \cref{fig:massive_example}. It is not clear at this point whether a designer potential can exist which allows for a region of $1+w_0 > 10^{-3}$, as we would naturally expect such a potential to also violate constraints on the fine-structure constant from the CMB or laboratory data. This confirms (and numerically strengthens) the analysis of \cite{Martins2016}, which derived a generic constraint of $1+w_0 < 10^{-7}$. We note that this bound is much tighter than generic CMB bounds on dynamical dark energy (with some $w_0 \neq -1$) at the approximate level of $1+w_0 < 10^0$ \cite{PlanckVI2020}; even when combined with BAO+supernovae, this only decreases to a level of $1+w_0 < 10^{-1}$ \cite{Schoneberg2022} -- this is because the bound is primarily driven by the fine structure constraint from the CMB (for the \enquote{Cosmo} case) or the fine-structure data (for the \enquote{All} case), as elucidated in the discussion in \cref{sec:discussion}.

Thus, the detection of rolling tachyon fields is not to be expected from cosmological data and remains only to be searched for in future fine-structure constant and weak equivalence principle measurements. A significant increase in the precision at which $\alpha$ is measured is foreseen both from laboratory measurements through the use of novel techniques such as nuclear clocks systems \cite{Beeks2024}, and from astrophysical observations of QSO at high redshifts using high precision spectrometers as ESPRESSO \cite{Pepe:2020ent} or ANDES \cite{Martins2024}. Such measurements represent our last lever arm in order to reveal -- or further constrain -- the possible existence of a rolling tachyon field in our Universe. Such an improvement in the precision of future measurements will most likely tighten the upper bounds on $1+w_0$\,, so that this model becomes even more indistinguishable from $\Lambda$CDM in terms of its dynamics. Even the observation of a drift in $\alpha$ consistent with a rolling tachyon model would not necessarily be distinguishable between it and another model generating such a drift. Therefore, future measurements are unlikely to increase our understanding of the model beyond tightening the bounds on its dynamics. As such, we find that the rolling tachyon can play at most a test field role in the cosmological evolution and does not show any significant non-trivial dynamics.

\section*{Acknowledgments}

\begin{sloppypar}
This work was financed by Portuguese funds through FCT (Funda\c c\~ao para a Ci\^encia e a Tecnologia) in the framework of the project 2022.04048.PTDC (Phi in the Sky, DOI 10.54499/2022.04048.PTDC). 
JDFD is supported by an FCT fellowship, grant number \\\mbox{SFRH/BD/150990/2021}. 
CJM also acknowledges FCT and POCH/FSE (EC) support through Investigador FCT Contract 2021.01214.CEECIND/CP1658/CT0001 (DOI 10.54499/2021.01214.CEECIND/CP1658/CT0001). 
NS acknowledges the support of the following Maria de Maetzu fellowship grant: Esta publicaci\'on es parte de la ayuda \mbox{CEX2019-000918-M}, financiada por MCIN/AEI/10.13039/501100011033. 
NS also acknowledges support by MICINN grant number PID \mbox{2022-141125NB-I00}.
LV acknowledges partial support by the Italian Space Agency LiteBIRD Project (ASI Grants No. 2020-9-HH.0 and 2016-24-H.1-2018), as well as the RadioForegroundsPlus Project HORIZON-CL4-2023-SPACE-01, GA 101135036. 
This computational work was produced with the support of INCD funded by FCT and FEDER under the Advanced Computing Project 2023.10382.CPCA.A1.
\end{sloppypar}

\bibliographystyle{JHEP}
\bibliography{biblio.bib}

\appendix
\section{Canonical scalar field limit}\label{app:canonical}

As noted in the main body of this work, there is a strong reason to investigate the regime of $\dot{\phi}\ll1$, which is very typical for this kind of model. One important consequence of such a bound is that we naturally find an equivalence to a typical canonical scalar field with a special coupling to the fine-structure constant. Taylor expanding the square-root in the Tachyon's Lagrangian density, we get $\mathcal{L} \simeq V\left(\phi\right)\left(1-\frac{1}{2}\dot{\phi}^2\right)$. Now, re-normalizing the field as $\tilde{\phi} = \int \sqrt{V} d\phi$ and realizing that $\dot{\phi} = V^{-1/2} \dot{\tilde{\phi}}$, one can write the Lagrangian density as 
\begin{equation}
    \mathcal{L} = \frac{1}{2}\dot{\tilde{\phi}}^2 - V\left(\phi\right) = \frac{1}{2}\dot{\tilde{\phi}}^2 - \tilde{V}\left(\tilde{\phi}\right)~,
\end{equation}
which is the equivalent form of a canonical scalar field with a different potential $\tilde{V}(\tilde{\phi})$ (we have put a tilde on the potential $V$ to stress this different form). We also note in passing that the field $\tilde{\phi}$ is unitless (as opposed to $\phi$).

An example for a canonical field would be for the massive potential that $\tilde{\phi} = \sqrt{V_0}/M \cdot (-i) \sqrt{\pi} \mathrm{erf}(i M \phi/2)$, and we observe that it is essentially a non-linear transformation of the $\eta$ we defined in that case.

This argument can also be made more rigorously with the full form of the DBI action incorporating the coupling of the tachyon with the gauge fields according to \cite{Garousi2005}
\begin{equation}
    S = -\int d^4x V\left(\phi\right)\sqrt{-\det\left(g_{\mu\nu} + \partial_\mu\phi\partial_\nu\phi + 2\pi\alpha^{\prime}\beta^{-1}F_{\mu\nu}\right)}~,
\end{equation}
where $\alpha'=1/M_s$ is the universal Regge slope (with string mass scale $M_s$ ) and $\beta$ is a warped factor. We can show that this can be viewed as a special coupling of the electromagnetic field tensor by simply expanding the expression in this field tensor. Letting $A_{\mu\nu} = g_{\mu\nu} + \partial_\mu\phi \partial_\nu \phi$ and $B_{\mu\nu} = 2\pi \alpha' \beta^{-1} F_{\mu \nu}$\,, we can first note generally that
\begin{equation}
\begin{aligned} \label{app:eq:determinant_expansion}
    \det (A+B) & = \sqrt{\det A} + \frac{\sqrt{\det A}}{2} \Tr (A^{-1}B)  \\ & - \frac{\sqrt{\det A}}{8} \left(2\Tr(A^{-1} B A^{-1} B)-\Tr(A^{-1}B)^2)\right) +\mathcal{O}(B^3)~,
\end{aligned}
\end{equation}
and we also immediately notice that $A_{\mu\nu} = A_{\nu\mu}$ and that $B_{\mu\nu} = - B_{\nu\mu}$. To get rid of the terms involving  $\Tr(A^{-1}B)$, we can use the property that the inverse of $A$ must also be totally symmetric, since
\begin{equation}
\begin{aligned}
     (A^{-1})_{\mu \sigma} &= I_{\mu \rho} (A^{-1})_{\rho \sigma}= I_{\rho \mu} (A^{-1})_{\rho \sigma} = A_{\rho \nu} (A^{-1})_{\nu \mu} (A^{-1})_{\rho \sigma}\\ &= A_{\nu\rho}(A^{-1})_{\rho \sigma} (A^{-1})_{\nu \mu}  =  I_{\nu \sigma} (A^{-1})_{\nu \mu} =  I_{\sigma \nu} (A^{-1})_{\nu \mu}= (A^{-1})_{\sigma \mu}~,
\end{aligned}
\end{equation}
where $I$ is the (symmetric) identity matrix and we have used the symmetry of $A$ in the line-break of the equation. The rest is relatively straightforward, since
\begin{equation}
    \Tr (A^{-1}B) = (A^{-1}B)_{\mu \mu} = A^{-1}_{\mu \nu} B_{\nu \mu} = -A^{-1}_{\nu \mu} B_{\mu \nu} = -(A^{-1}B)_{\nu \nu} = - \Tr(A^{-1}B)~,
\end{equation}
and therefore $\Tr(A^{-1}B)=0$. Collecting the non-zero terms from \cref{app:eq:determinant_expansion} we see that
\begin{equation}
    \sqrt{\det(A+B)} = \sqrt{\det A} + \frac{\sqrt{\det A}}{4} \Tr( A^{-1} B A^{-1} B)~.
\end{equation}
Re-expressing $A$ and $B$ one finds at leading order
\begin{equation}
    S\simeq - \int d^4x \left[ V\left(\phi\right)\sqrt{-\det\left(g_{\mu\nu} + \partial_\mu\phi\partial_\nu\phi \right)} + \frac{\left(2\pi\alpha^{\prime}\right)^2V\left(\phi\right)}{4\beta^2}\sqrt{-g}\Tr\left(g^{-1}Fg^{-1}F\right) \right]~,
\end{equation}
which can be further simplified using
\begin{equation}
    \Tr\left(g^{-1}Fg^{-1}F\right) = g^{\mu\rho}F_{\rho\sigma}g^{\nu\sigma}F_{\mu\nu} = F^{\mu \nu}F_{\mu \nu}~,
\end{equation}
which allows us to make the comparison to the canonical coupling of the form $\frac{\sqrt{-g}}{4}B_F(\phi)F^{\mu\nu} F_{\mu\nu}$ with a special coupling of the form
\begin{equation}\label{eq:BF}
  B_F\left(\phi\right)\equiv\left(\frac{2\pi}{\beta M_s^2}\right)^2 V\left(\phi\right)~,
\end{equation}
where the consistency relation to achieve Maxwellian dynamics in the electromagnetic sector in laboratory experiments today implies that we must assume that
\begin{equation}
    \left(\frac{2\pi}{\beta M_s^2}\right)^2 V(\phi_0)=1 \quad \Rightarrow \quad \beta = \frac{2\pi}{M_s^2} \sqrt{V(\phi_0)}~,
\end{equation}
see also for example \cite[eq.~(24)]{Garousi2005}. With that choice, we can simplify the expression and obtain
\begin{equation}
B_F\left(\phi\right)\equiv\frac{V\left(\phi\right)}{V(\phi_0)}~.
\end{equation}

We also note that in the canonical field limit \cref{eq:BF} becomes simply 
\begin{equation}
  B_F\left(\tilde{\phi}\right)\equiv \frac{\tilde{V}\left(\tilde{\phi}\right)}{\tilde{V}\left(\tilde{\phi_0}\right)}~.
\end{equation}

\section{Constraints from the E\"otv\"os parameter \label{app:eotvos}}

Given the discussion in \cref{app:canonical}, we can now proceed to estimate roughly what kind of constraining power we should be expecting from experiments involving the freefall of test masses. This is because any change in the fine-structure constant induces changes in the binding energy, therefore effectively violating the weak equivalence principle.\footnote{A simple explanation can be found in \cite[appendix B]{Schoneberg:2024ynd}. Additionally, explicit derivations with many more details can be found for example in \cite{Chiba2002,Dvali2002}, which we summarize here: Since the rolling tachyon imparts a variation of the fine-structure constant, such a variation will effectively cause different binding energies of the atoms, depending on the value of the tachyon field at the given location. More precisely, the atomic constituents can effectively exchange virtual tachyons that generate a kind of fifth force, leading to an acceleration that can be detected.} In laboratory experiments the differential acceleration between different test particles with different binding energies could in principle be measured. Currently there is no preference of a deviation from universal free fall, meaning that the differential acceleration is restricted at the order of $\eta_\text{E\"otv\"os} \sim 10^{-15}$ \cite{Touboul2022}.

This acceleration has also been predicted for canonical scalar fields, where arguments from \cite{Chiba2002} have shown it to be of the order of\footnote{The prefactor is not precisely determined, usually values between $10^{-3}$ and $10^{-4}$ are cited, see also \cite{Dvali2002,Chiba2002}.} We decided to remain conservative here, as a value of $10^{-3}$ would lead to an even tighter bound of $\kappa^2/H_0 \sim 10^{-12}$.
\begin{equation}
    \eta_\text{E\"otv\"os} \sim 10^{-4} \lambda^2~,
\end{equation}
where $\lambda$ is the derivative of the coupling to the electromagnetic field (above $B_F(\phi)$) for such a canonical scalar field. In order to apply such a bound to the rolling tachyon field, a few comments are required. In the original works (see \cite{Chiba2002}), the potential necessarily involves a mass-term, which used to derive the corresponding Feynman amplitudes and a Yukawa-like interaction. This is not always possible for the rolling tachyon field, especially in potentials other than the massive one. Additionally, the original works assume standard canonical Lagrangians. As discussed in \cref{app:canonical}, such a limit is possible with the tachyon field (given that $\dot{\phi} \ll 1$, which we have argued for in \cref{sec:discussion}), but with a slightly modified field $\mathrm{d}\tilde{\phi} = \mathrm{d}{\phi} \sqrt{V}$. Assuming, for the moment, that the same calculations as in \cite{Dvali2002} would still apply (with possibly modified non-Yukawa effective potentials that still have the same Coulomb-like limit for the correct order of magnitude of the tachyon potential parameters), we would find that
\begin{equation}
    10^{-5.5} \sim \lambda = -\frac{\mathrm{d} B_F(\tilde{\phi})}{\mathrm{d}\tilde{\phi}}\bigg|_{z=0} = - \frac{\mathrm{d}\ln \tilde{V}(\tilde{\phi})}{\mathrm{d}\tilde{\phi}}\bigg|_{z=0} = \frac{\mathrm{d}\ln V(\phi)}{\mathrm{d}\phi} \frac{1}{\sqrt{V_0}}~,
\end{equation}
where we have explicitly used the canonical field $\tilde{\phi}$ instead, and used its defining property. Now we can estimate the approximate order of magnitude of this constraint using a similar argument as in \cref{sec:discussion} (therefore using $\partial_\phi \ln V \approx \kappa$) as well as explicitly using the fact that $V_0 \sim H_0^2$ due to the late-time dark energy domination (and $|1+w_0| \ll 1$) to find
\begin{equation}
    10^{-5.5} \sim \lambda = \frac{\mathrm{d}\ln V(\phi)}{\mathrm{d}\phi} \frac{1}{\sqrt{V_0}} \sim \kappa/H_0~.
\end{equation}
This would allow us to conclude that \emph{under the aforementioned approximation} that the E\"otv\"os free-fall constraint would lead to constraints on $\kappa^2/H_0^2 \sim 10^{-11}$, to be compared with the fine-structure constraint of the order $\kappa^2/H_0^2 \sim 10^{-9}$ and the cosmological constraints of the order $\kappa^2/H_0^2 \sim 10^{-3}$. We therefore expect such free-falling constraints to potentially have a large impact, though we suspect that the entire underlying theory of the fifth-force approximation discussed in \cite{Chiba2002,Dvali2002} would have to be explicitly be generalized for this case, and we leave this for future work.

\section{Cosmological corner plots}\label{app:triangles}

In this appendix we display the triangle corner plots for the cosmological parameters (like $H_0, n_s, \ldots$) in all of the different investigated potentials. We don't notice any strong deviations of the parameter constraints from $\Lambda$CDM, other than a very mild allowance for slightly smaller $H_0$ values in some of the models. For the potentials where two different regions of potential parameter space are explored these corner plots don't change significantly, so we just present here just one of them.

\begin{figure}[H]
\centering
\includegraphics[width=0.85\textwidth]{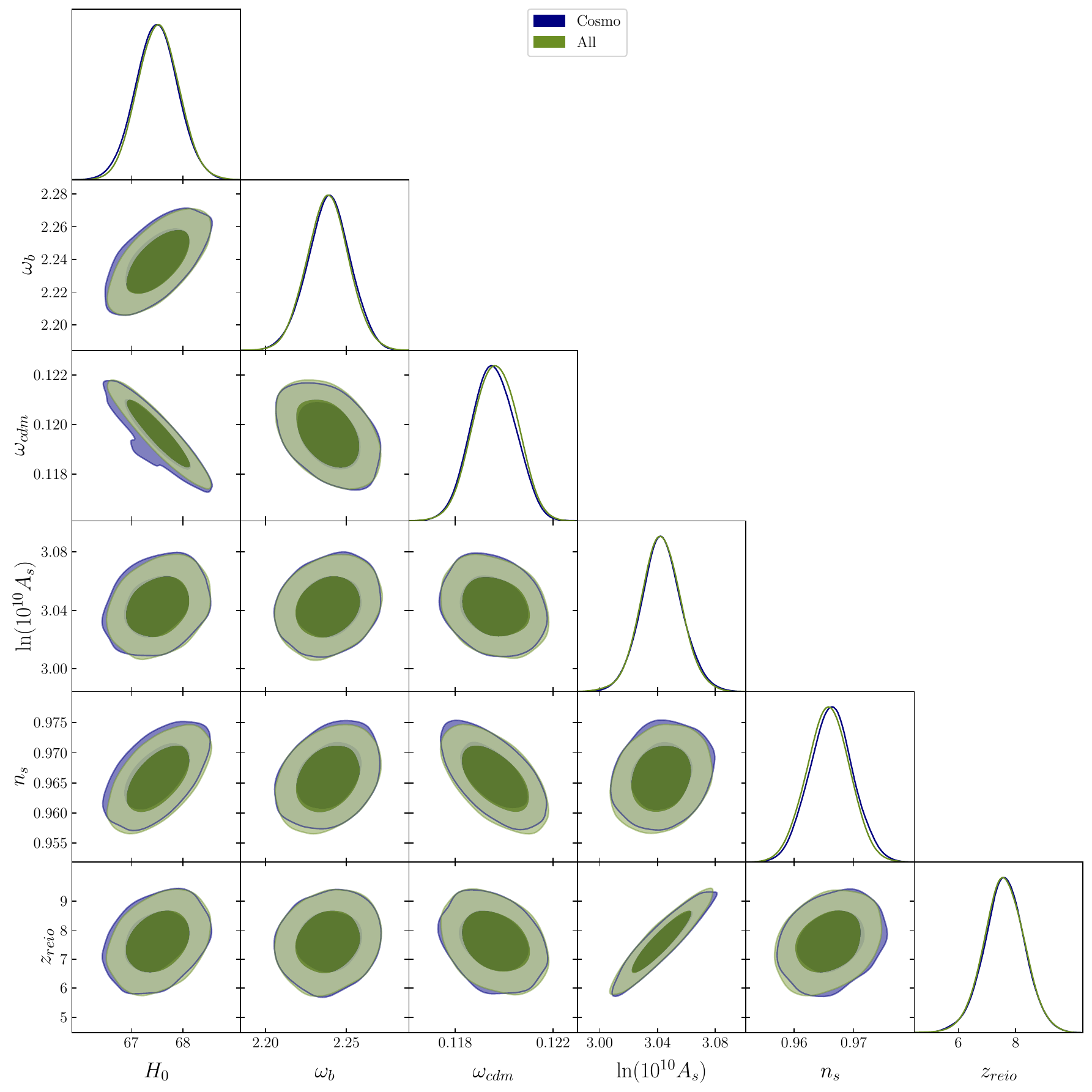}
\caption{Same as \cref{fig:massive_alpha} (massive potential), but for the cosmological parameters instead.}
    \label{fig:massive_cosmo}
\end{figure}
\begin{figure}[H]
\centering
\includegraphics[width=0.85\textwidth]{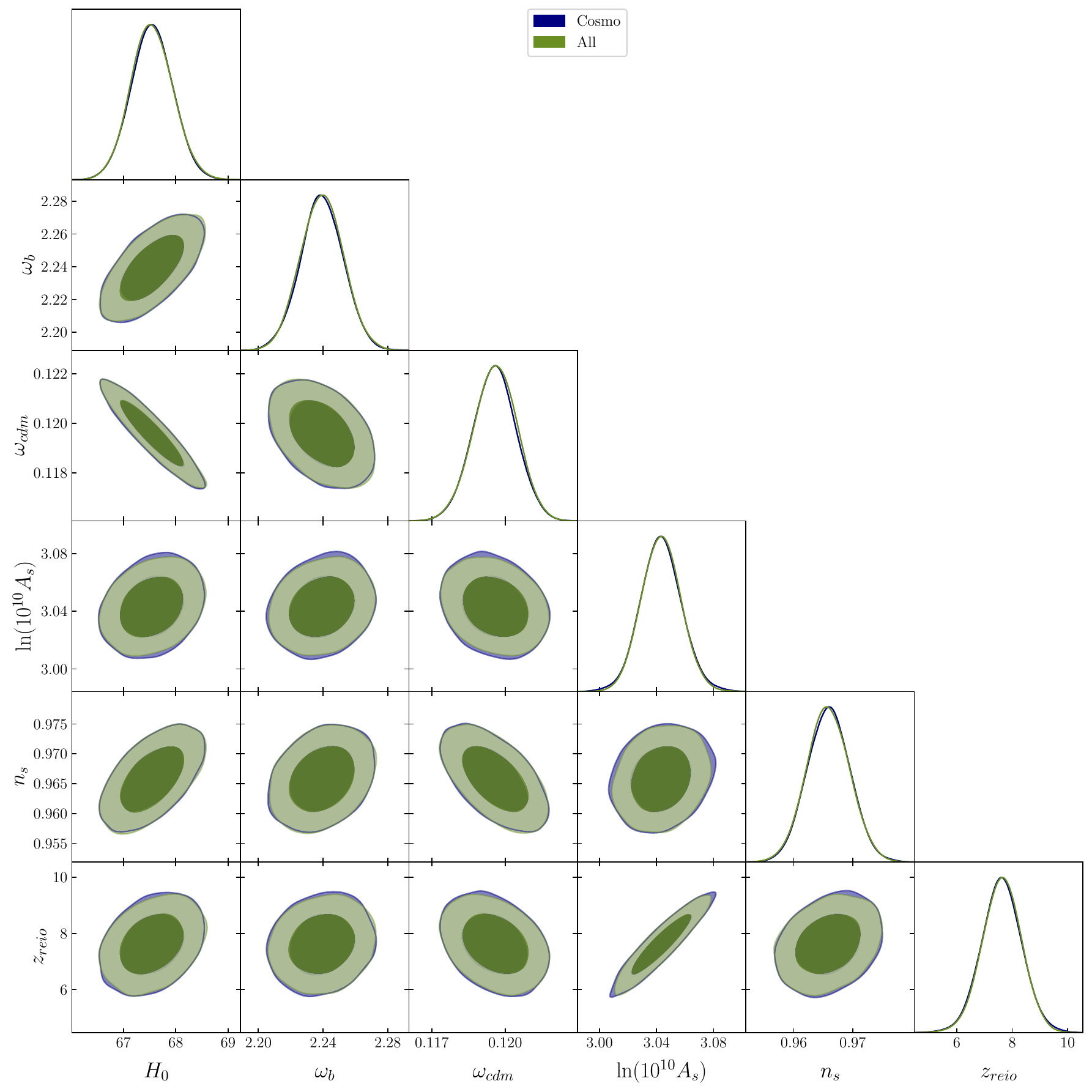}
\caption{Same as \cref{fig:massive_cosmo}, but for the inverse power-law potential.}
    \label{fig:invpow_cosmo}
\end{figure}
\begin{figure}[H]
\centering
\includegraphics[width=0.85\textwidth]{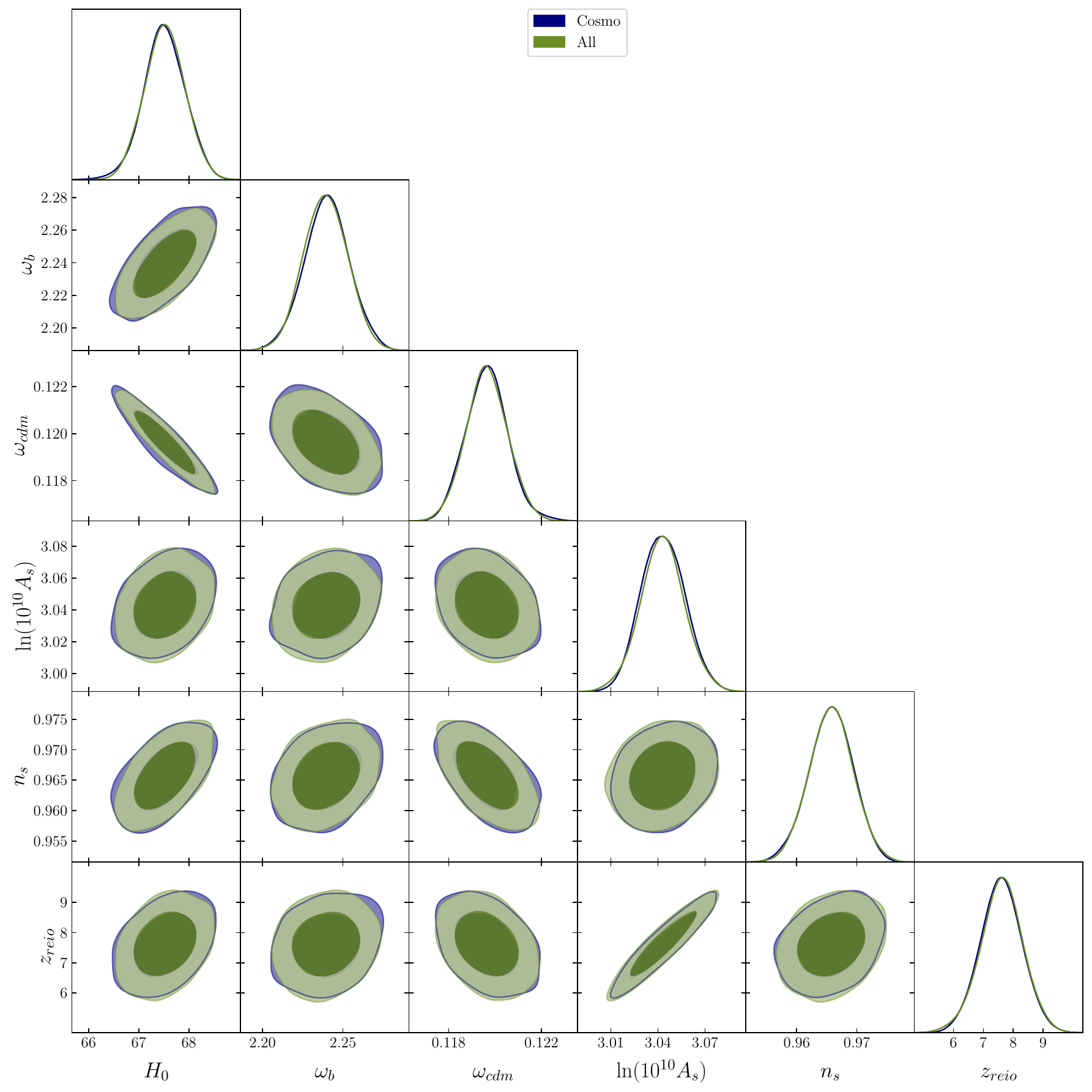}
\caption{Same as \cref{fig:massive_cosmo}, but for the anti-massive potential.}
    \label{fig:anti_cosmo}
\end{figure}
\begin{figure}[H]
\centering
\includegraphics[width=0.85\textwidth]{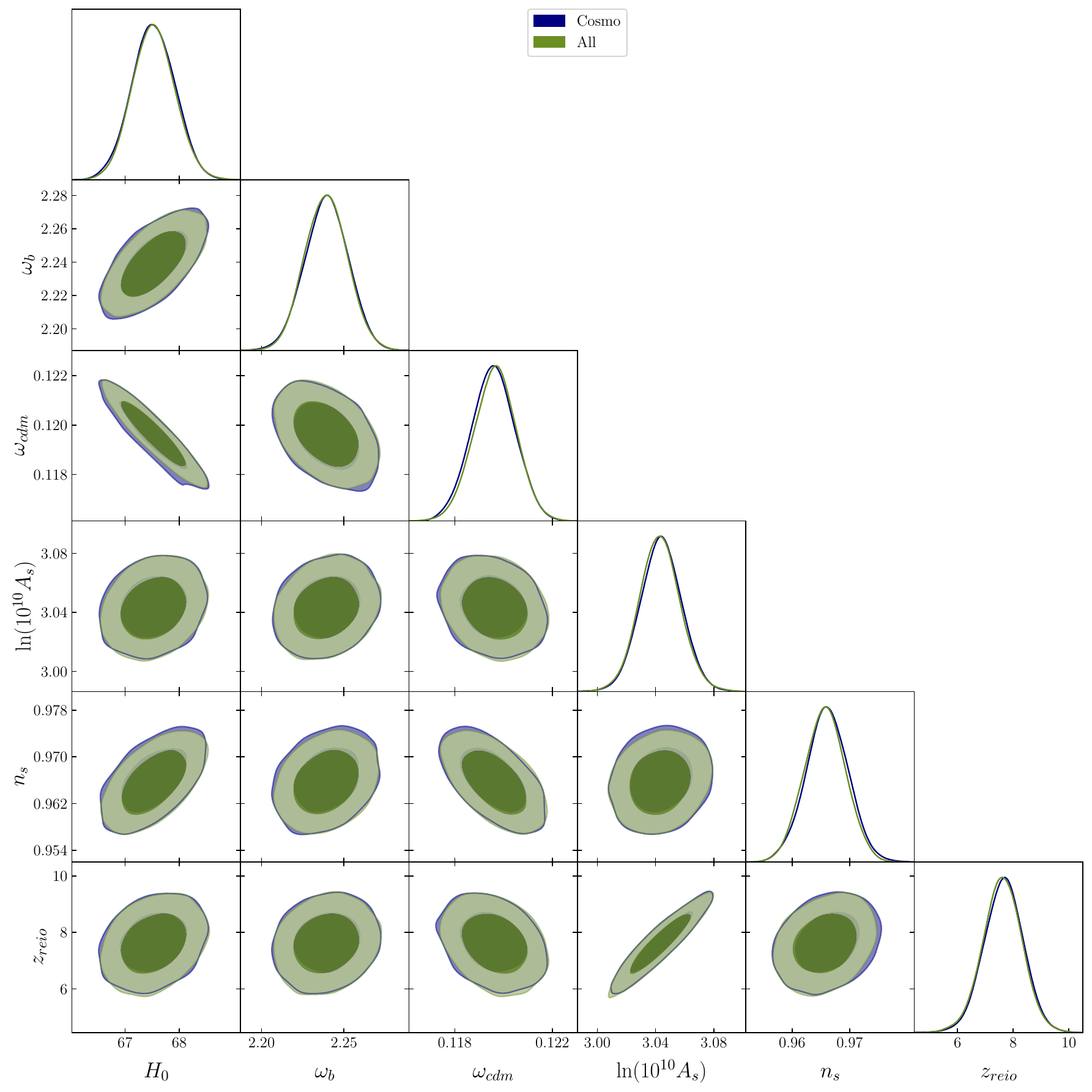}
\caption{Same as \cref{fig:massive_cosmo}, but for the negative exponential potential.}
    \label{fig:nega_cosmo}
\end{figure}
\begin{figure}[H]
\centering
\includegraphics[width=0.85\textwidth]{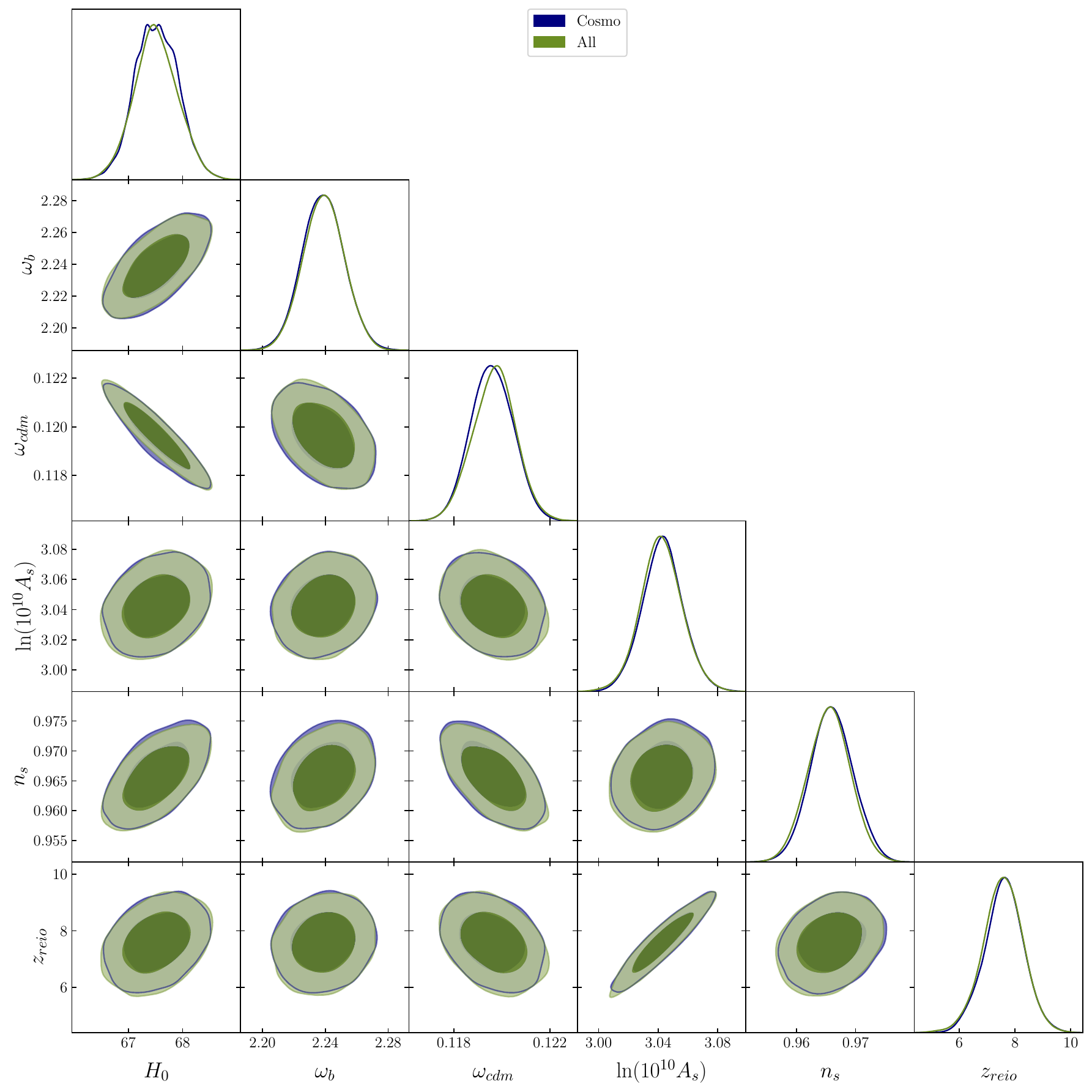}
\caption{Same as \cref{fig:massive_cosmo}, but for the inverse exponential potential.}
    \label{fig:invexp_cosmo}
\end{figure}
\begin{figure}[H]
\centering
\includegraphics[width=0.75\textwidth]{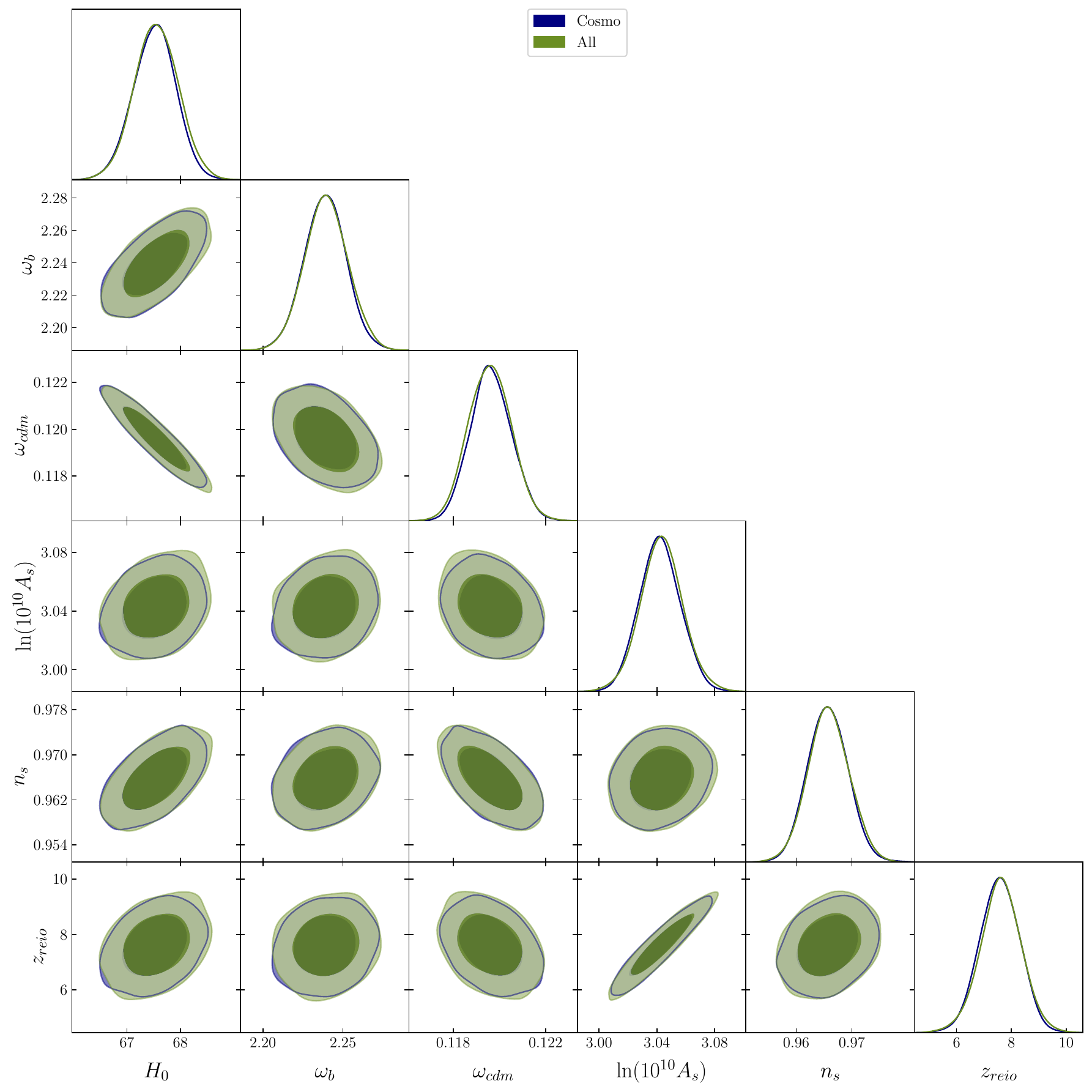}
\caption{Same as \cref{fig:massive_cosmo}, but for the hyperbolic cosine potential.}
    \label{fig:cosh_cosmo}
\end{figure}

\end{document}